\documentclass[final]{agujournal2019}
\usepackage{url} 
\usepackage[thicklines]{cancel}
\usepackage{soul}
\usepackage[version=4]{mhchem}
\journalname{JGR: Planets}

\begin{document}

\title{Growth and Evolution of Secondary Volcanic Atmospheres: II. The Importance of Kinetics}

%
%

\authors{Philippa Liggins\affil{1}, Sean Jordan\affil{2}, Paul B. Rimmer\affil{1}\affil{3}\affil{4}, Oliver Shorttle\affil{1}\affil{2}}

\affiliation{1}{Department of Earth Sciences, University of Cambridge, United Kingdom} 
\affiliation{2}{Institute of Astronomy, University of Cambridge, United Kingdom} 
\affiliation{3}{Cavendish Laboratory, University of Cambridge, United Kingdom} 
\affiliation{4}{MRC Laboratory of Molecular Biology, Francis Crick Ave, Cambridge, United Kingdom} 


\correspondingauthor{Philippa Liggins}{pkl28@cam.ac.uk}



\begin{keypoints}
\item Thermochemical equilibrium cannot be assumed for volcanically-derived atmospheres with temperatures $<$\,700K
\item Quenching of volcanically-derived atmospheres will limit linking of their chemistry to underlying mantle $f\mathrm{O_2}$
\item Warm to cool volcanic atmospheres produce \ce{CO2} and \ce{CH4}, but with CO present, preventing their being mistaken for positive biosignatures
\end{keypoints}

%
%

\begin{abstract}
	Volcanism is a major and long-term source of volatile elements such as C and H to Earth's atmosphere, likely has been to Venus's atmosphere, and may be for exoplanets. Models simulating volcanic growth of atmospheres often make one of two assumptions: either that atmospheric speciation is set by the high-temperature equilibrium of volcanism; or, that volcanic gases thermochemically re-equilibrate to the new, lower, temperature of the surface environment. In the latter case it has been suggested that volcanic atmospheres may create biosignature false positives. Here, we test the assumptions underlying such inferences by performing chemical kinetic calculations to estimate the relaxation timescale of volcanically-derived atmospheres to thermochemical equilibrium, in a simple 0D atmosphere neglecting photochemistry and reaction catalysis. We demonstrate that for planets with volcanic atmospheres, thermochemical equilibrium over geological timescales can only be assumed if the atmospheric temperature is above $\sim$700\,K. Slow chemical kinetics at lower temperatures inhibit the relaxation of redox-sensitive species to low-temperature thermochemical equilibrium, precluding the production of two independent biosignatures through thermochemistry alone: 1. ammonia, and 2. the co-occurrence of \ce{CO2} and \ce{CH4} in an atmosphere in the absence of CO. This supports the use of both biosignatures for detecting life. Quenched at the high temperature of their degassing, volcanic gases also have speciations characteristic of those produced from a more oxidized mantle, if interpreted as being at thermochemical equilibrium. This therefore complicates linking atmospheres to the interiors of rocky exoplanets, even when their atmospheres are purely volcanic in origin. 
\end{abstract}

\section*{Plain Language Summary}
Rocky planets can build up atmospheres over time through the release of volcanic gases. Simulations of this process usually assume that the chemistry of these atmospheres will either be controlled by the temperature the gases were erupted at, or by the current temperature of the atmosphere. We test these assumptions by calculating the time it will take for the chemistry of an atmosphere built of volcanic gases to change from being controlled by the temperature of eruption, to a chemistry reflecting the current atmospheric temperature. We find that without additional processes (e.g., atmospheric photochemistry or reaction catalysis) speeding up the rates of reactions, atmospheres with temperatures below 700K will always have chemistries which reflect their emission temperature, rather than the current atmospheric temperature. Cool planets with volcanically-derived atmospheres should therefore not be modelled while assuming the atmospheric chemistry is controlled by the current temperature. These results also support the use of both ammonia and the combined presence of carbon dioxide and methane (in the absence of carbon monoxide) as biosignatures for detecting the presence of life on other planets.

%
%

\section{Introduction}

Exoplanet science will provide many fascinating insights into planet formation and evolution, one of which will be a greater understanding of the origin and nature of rocky planet atmospheres -- a possibility that recently became a lot closer to being realised with the launch of the James Webb Space Telescope \cite{lustig-yaeger2019DetectabilityCharacterizationTRAPPIST1}. Of particular importance for linking to planetary habitability, and in the search for life, is developing an understanding of how atmospheric observations of exoplanets might be used to infer their geological properties and processes. As such, many models of the atmospheres of rocky exoplanets have been developed to allow future observations to be interpreted in the context of different geological paradigms: magma ocean worlds \cite{elkins-tantonLinkedMagmaOcean2008, hirschmann2012MagmaOceanInfluence, hamano2013EmergenceTwoTypes, hamano2015lifetimespectralevolution, katyal2020EffectMantleOxidation, sossi2020RedoxStateEarth, lichtenberg2021VerticallyResolvedMagma, gaillard2022RedoxControlsMagma}; stagnant lid tectonics \cite<e.g.,>[]{noack2017VolcanismOutgassingStagnantlid, tosi2017HabitabilityStagnantlidEarth, dorn2018OutgassingStagnantlidSuperEarths}; impact events \cite{schaefer2010ChemistryAtmospheresFormed, schlichting2015AtmosphericMassLoss, sinclair2020EvolutionEarthAtmosphere, zahnle2020CreationEvolutionImpactgenerated, kuwahara2015MolecularCompositionImpactgenerated}; interior structure and chemistry \cite{noack2014CanInteriorStructurea, schaefer2017RedoxStatesInitial, dorn2018OutgassingStagnantlidSuperEarths, ortenzi2020MantleRedoxState, spaargaren2020InfluenceBulkCompositiona, abe2011ProtoatmospheresSurfaceEnvironment}; surface chemistry \cite{schaefer2012VaporizationEarthApplication, herbort2020AtmospheresRockyExoplanets, yu2021HowIdentifyExoplanet}; atmospheric escape \cite{wordsworth2014ABIOTICOXYGENDOMINATEDATMOSPHERES, kislyakova2015StellarDrivenEvolution, dong2018AtmosphericEscapeTRAPPIST1, kite2020ExoplanetSecondaryAtmosphere, tian2015HistoryWaterLoss}; or volcanism \cite{gaillard2014TheoreticalFrameworkVolcanic, liggins2020CanVolcanismBuild, wogan2020AbundantAtmosphericMethane, kadoya2015EVOLUTIONARYCLIMATETRACKS}.

This paper series explores an important end-member case of hot rocky planet atmospheric evolution, where their atmospheres have been formed and evolved purely under the influence of volcanic gas supply: i.e., without the oceans, aqueous geochemistry or life that exert strong controls on Earth's atmosphere. As has already been explored in Paper I \cite{liggins2022GrowthEvolutionSecondary}, these planets are of particular interest because they can form classes of volcanic atmosphere which are linked to the interior conditions (particularly, how oxidised planetary mantles are) which may reflect core formation history, planetary mass, and formation location within the protoplanetary disk \cite{wade2005CoreFormationOxidation, frost2008RedoxStateMantle}. Paper III will expand on the results in Paper I by including considerations of atmospheric escape. These end-member models will be important to inform evidence of volcanic activity in future atmospheric abundance estimates made for exoplanet spectra.

A key question in making this link between a planet's atmosphere and interior is to what extent the lower pressures and temperatures of a planet's atmosphere will allow it's chemistry to reflect information about the geochemistry of its interior. A commonly used approximation for modelling rocky planet atmospheres is to assume that they will exist in thermochemical equilibrium, either homogeneous \cite<equilibrium within the gas phase only, e.g.,>[]{liggins2020CanVolcanismBuild, ortenzi2020MantleRedoxState, wogan2020AbundantAtmosphericMethane} or heterogeneous \cite<equilibrium is extended to include surface interactions, e.g.,>[]{herbort2020AtmospheresRockyExoplanets, sossi2020RedoxStateEarth, woitke2021CoexistenceCH4CO2}. However, in volcanically-derived atmospheres, gases will typically be injected into the atmosphere at higher temperatures than the ambient surface temperature. The extent to which these high temperature volcanic gases will relax and re-equilibrate to a new, low temperature equilibrium is unclear, but is important for understanding how interior conditions may be inferred from an atmosphere. For example in Paper I \cite{liggins2022GrowthEvolutionSecondary}, we found that clear atmospheric classes (defined according to the speciation of the atmosphere at thermochemical equilibrium) can emerge in volcanically--derived atmospheres as a function of mantle $f\mathrm{O}_2$. However, while these classes form most clearly under lower atmospheric temperature conditions, they require thermochemical equilibrium to have been reached in order to be valid.

Thermochemical equilibrium in low temperature atmospheres derived by outgassing of silicate materials has previously been invoked as a potential source of false-positive biosignatures. \citeA{woitke2021CoexistenceCH4CO2} found that at thermochemical equilibrium with temperatures $\leq$\,600\,K, gases released by ``common rock materials" (carbonaceous chondrites and mid-ocean ridge basalt) produced an atmosphere where \ce{CO2} and \ce{CH4} could coexist with only trace amounts of CO, producing a geological false-positive for the \ce{CO2}+\ce{CH4} biosignature of \citeA{krissansen-totton2018DisequilibriumBiosignaturesEarth}. However, this conclusion depends heavily on the timescale over which a low temperature equilibrium can be achieved.

In a system modelled at thermochemical equilibrium (as would be the case for calculations performing Gibbs-free energy minimisation), the timescale for gas chemistry to adjust to the surface P-T conditions is necessarily just the cooling rate of the gas. The (sometimes implicit) assumption here is that reaction rates are fast enough to adjust speciation to the new equilibria. However, as the temperature and pressure of a volcanic gas decreases towards ambient surface conditions, both the frequency of molecular collisions, and the energies with which these collisions occur, decrease. The timescale over which a new equilibrium can be achieved thus lengthens. The consequence of this is that volcanic gases as observed on Earth, which are erupted at high temperatures into a cold environment and cool rapidly, usually have their chemistry `quenched' \cite{leguern1982FieldGasChromatograph, gerlach1986EvaluationGasData}: their gas-phase speciation is frozen in at the temperature where chemical reactions become slow enough that the speciation of the gas is effectively constant over the timescale of interest. In detail, this quench temperature will vary by species, so that the resultant quenched gas entering and mixing with the atmosphere reflects the equilibrium state of no single temperature (or pressure).

In defining quenching to have occurred, the timescale of interest is critical. If it takes a thousand years for a volcanic gas to re-equilibrate at a cooler temperature, it may be considered quenched over the timescale of observations on Earth (usually minutes to a few days while the gas mixes into the background atmosphere), but not when looking at gas chemistry over geological time (i.e., on the order of billions of years). Depending on the rate at which a cooled volcanic gas re-attains thermochemical equilibrium, volcanically-derived atmospheres of rocky planets may be either in thermochemical equilibrium, as many modelling results assume, or a disequilibrium gas mixture, where volcanic gases are present in a quenched composition from some higher temperature than the background atmosphere.

Here, we identify the temperature below which chemical equilibrium is no longer a valid assumption, and use this to comment on the implications for various approaches to both modelling volcanic atmospheres, and applying potential biosignatures. In Sect.\,\ref{section:methods}, we present a brief recap of our EVolve model used to produce synthetic atmospheres, which is described in full in Paper I, along with our methods for calculating chemical timescales. Results are shown in Sect.\,\ref{section:results}, including a description of the rate limiting steps we identify as preventing the chemical relaxation of volcanic atmospheres (Section\,\ref{section:rates}). The discussion includes an assessment of model limitations, and a demonstration that the key results of this study are independent of the chemical network chosen to perform the modelling (see Section\,\ref{section:minimum_temps}). Conclusions follow in Sect.\,\ref{section:concs}.

\section{Methods}
\label{section:methods}

The conceptual model applied here is similar to those used in previous studies which assume thermochemical equilibrium in an atmosphere; we model the thick, near-surface layer of the atmosphere, conceptually a 0D box at a single (surface) temperature and pressure. This box ignores surface-atmosphere interactions outside of input by volcanic outgassing. We also ignore the effects of vertical mixing and photochemistry within the atmosphere, both of which would introduce disequilibrium into the near-surface environment.

\subsection{Evolving volcanic atmospheres}
The evolution and chemical composition of volcanically--derived atmospheres in thermochemical equilibrium is modelled here using the EVolve atmospheric evolution model, previously described in Paper I \cite{liggins2022GrowthEvolutionSecondary}. In brief, EVolve is a 3-part model linking the mantle to the atmosphere using EVo \cite{liggins2022EVO_rtd}, a volcanic degassing model for elements C, O, H, S and N. In every time-step, a portion of the mantle is melted and volatiles partition from the bulk mantle into the melt phase according to the batch melting equation. The mass and volatile content of this magma, along with how oxidising the planet's upper mantle is (measured by its oxygen fugacity, the $f\mathrm{O}_2$ relative to the iron-w{\"u}stite, IW, rock buffer) are used as initial conditions in EVo. EVo finds the volatile saturation pressure of this magma, then calculates the volatile element chemistry of the two-phase magma and exsolved gas mixture at both the saturation pressure, and the surface at a constant temperature of 1200\textdegree{}C. This is done by simultaneously solving a system of 5-7 heterogeneous equilibria (between silicate melt and gas, described using solubility laws), 5 homogeneous gas-phase equilibria and the balance between oxygen stored in the melt as \ce{Fe2O3} or \ce{FeO}, and oxygen within the volatile species. The end point of the volcanic outgassing is defined as the surface atmospheric pressure, which gradually increases with every time-step as volatile mass is added to the atmosphere.

The FastChem 2.0 equilibrium chemistry model \cite{stock2018FastChemComputerProgram, stock2021} is next used to calculate the chemistry of the atmosphere at the recalculated surface pressure and temperature of the planet for the new, mixed atmosphere. It is this 0D atmospheric temperature which is varied to determine the timescales to equilibrium for different species, while keeping the temperature of the volcanic melt+gas system constant. We note that FastChem 2.0 assumes an ideal gas equation of state, but with the maximum pressures investigated in this study of $\le30\,$bar little deviation between fugacities and partial pressures are expected.

The results presented in this paper are all for a single set of initial mantle volatile contents, with a \ce{H2O}/\ce{CO2} mass ratio of 9 and a C:S:N ratio similar to modern MORB magmas on Earth \cite<See Paper I,>[]{liggins2022GrowthEvolutionSecondary}. This results in an initial mantle volatile content of 450\,ppm H$_2$O, 50\,ppm CO$_2$, 54\,ppm S, and 0.36\,ppm N. Results are always presented relative to a mantle $f\mathrm{O}_2$, reflecting the redox state in the mantle at the pressure of melt production. The final $f\mathrm{O}_2$ of a melt+gas parcel will change according to the degree of volatile outgassing which occurs \cite<for example, see>[]{carmichael1991RedoxStatesBasic,oppenheimer2011MantleSurfaceDegassing,brounce2017RedoxVariationsMauna} and will therefore vary through time as the volatile content of the melt and outgassing pressure changes; the constant $f\mathrm{O}_2$ of the mantle is therefore provided as a fixed independent reference point for a specific planet.

\subsection{Calculating chemical timescales}
To test how long it may take for a volcanic atmosphere to reach thermochemical equilibrium we used a chemical-kinetics code for planetary atmospheres.  This code consists of a solver, ARGO \cite{rimmerCHEMICALKINETICSNETWORK2016, rimmer2019ErratumChemicalKinetics}, and a chemical network, STAND2020 \cite{hobbs2021SulfurChemistryAtmospheres, rimmer2021HydroxideSaltsClouds}. The reactions listed in the chemical network are solved by ARGO as a set of time-dependent, coupled, nonlinear differential equations, with reaction rates calculated as functions of pressure, temperature, and concentrations of the reactant species. As the solver  proceeds, and the chemical composition of the atmosphere tends closer to its equilibrium speciation, the chemical timescales can become very long and, in certain cases, the solver subroutine can stall before chemical equilibrium is achieved. This is generally the result of the chemical speciation reaching a quasi-steady state, intermediate between the initial speciation and the equilibrium speciation. Stalling at quasi-steady state is a common problem in numerical modelling of chemical kinetics, which are typically stiff systems of differential equations \cite{leal2017KineticsStiffSystems}.

Where ARGO could not reach equilibrium, i.e., where it stalled with a quasi-steady state speciation different from that predicted for thermochemical equilibrium, instantaneous chemical timescales were calculated. The chemical timescales provide an estimate of the length of time it would take for each species to reach its abundance at chemical equilibrium, calculated as
\begin{linenomath*}
\begin{equation}
    \tau_{i,\,chem} = \frac{|n_{i}-n_{i,\,eq}|}{\frac{dn_{i}}{dt}},
\label{eq:timescale}
\end{equation}
\end{linenomath*}
where $\tau_{i,\,chem}$ (s) is the chemical timescale, $n_{i}$ (cm$^{-3}$) is the final number density of species $i$ calculated using ARGO, $n_{i,\,eq}$ (cm$^{-3}$) is the equilibrium number density of species $i$ calculated using FastChem 2.0 \cite{stock2018FastChemComputerProgram, stock2021}, and ${dn_{i}}/{dt}$ (cm$^{-3}$s$^{-1}$) is the rate of change of the number density of species $i$ as calculated by ARGO. For these simulations, ARGO was given as initial conditions: the equilibrium speciation of the gas at $1473\,{\rm K}$ (the temperature of the gas in the volcanic system) calculated using FastChem, the atmospheric temperature (between $500 - 2000\,{\rm K}$), and the pressure was set at the final pressure of the gas that would result after $1\,{\rm Gyr}$ of volcanic activity.

In all timescale calculations, the starting gas composition was taken as the speciation of a volcanically constructed atmosphere with a temperature of 1473\,K after 1\,Gyr of volcanic activity. The atmosphere was then assumed to be instantaneously cooled (or heated) to a new surface temperature, and ARGO was used to calculate the chemical speciation of the gas as a function of time as it adjusted towards the new thermochemical equilibrium. A discussion of how this relates to more realistic natural scenarios, and limitations of the approach, can be found in section \ref{section:caveats}.

As $n_i$ approaches $n_{i,\,eq}$, the rate of change of $n_i$ diminishes and equation \eqref{eq:timescale} approaches a singularity, causing the estimated chemical timescales to become very long. Equation \eqref{eq:timescale} is therefore only an appropriate estimate of the timescale to equilibrium if the system has not yet neared its equilibrium state. We therefore use eq.\,\eqref{eq:timescale} to estimate the timescale to equilibrium only for the systems that have not reached chemical equilibrium and the solver has instead stalled with some quasi-steady state abundance. For the systems that have reached chemical equilibrium we need only record the time taken for the solver to achieve this state. We can then compare the timescales to equilibrium, either explicitly solved for or estimated using eq.\,\eqref{eq:timescale}, to relevant geological timescales to determine whether or not a species will quench with a non-equilibrium abundance at a given atmospheric temperature. These calculations are performed for atmospheres with a temperature range $500 - 2000\,{\rm K}$. Examples of the kinetic evolution of the atmosphere are shown in \ref{section:relax_appendix}. In Fig.\,\ref{fig:relax_hot_warm_cool} we demonstrate the chemical relaxation in a `hot' (2000\,K), `warm' (1000\,K), and `cool' (500\,K) case, and we validate the instantaneous timescale calculation in cases where the timescales to true equilibrium were reached.

\section{Results}
\label{section:results}

\begin{figure}[ht]
    \centering
    \includegraphics[width=0.8\textwidth]{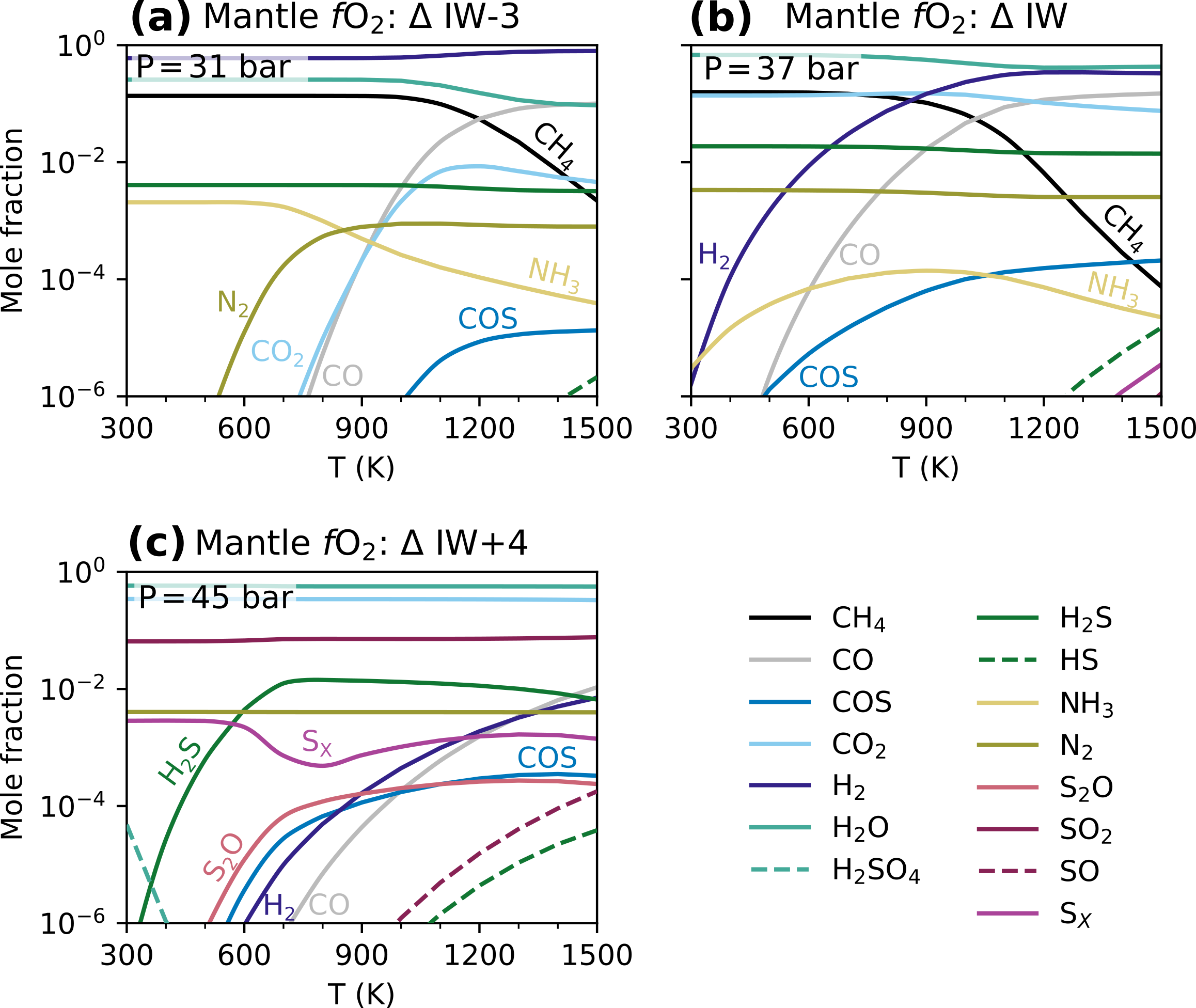}
    \caption{The chemical speciation of a volcanic atmosphere at the surface after 1\,Gyr of outgassing, shown for three values of the $f\mathrm{O}_2$ of the mantle supplying those gases and plotted as a function of the 0D atmospheric temperature. The surface pressure is labelled on each panel, as controlled by the mass of volatiles outgassed after 1\,Gyr. Thermochemical equilibrium is assumed, and other processes such as water condensation and photochemistry are ignored.}
    \label{fig:t_change}
\end{figure}

If planetary atmospheres are assumed to be in pure thermochemical equilibrium, then the speciation of gases in the atmosphere will change continuously with temperature (Fig.\,\ref{fig:t_change}; note that water condensation at low temperatures in these plots is ignored). The chemical speciation of the atmosphere tends to diminish in complexity as the temperature decreases, so that by 300\,K, regardless of the mantle $f\mathrm{O}_2$, the atmospheric speciation above the ppm level can be described by $\sim$ 5 species, as opposed to the $>$10 species which are present above ppm levels at 1500\,K. 

Two notable transitions in atmospheric composition during cooling are (i) the decrease in CO due to its hydrogenation to \ce{CH4}, and (ii) the decrease in \ce{N2} as it is hydrogenated to make \ce{NH3} (Figs.\,\ref{fig:t_change}a \& b). These changes in some of the major atmospheric species are key discriminators of different atmospheric classes, and a planet's mantle $f\mathrm{O}_2$ (see Paper I). However, the efficient operation of these reactions is strongly inhibited by low temperatures.

The extent to which \ce{CH4} and \ce{NH3} can be produced abiotically in planetary atmospheres has been a subject of debate over recent years \cite{seager2013BIOSIGNATUREGASESH2DOMINATED, schaefer2017RedoxStatesInitial,  wogan2020AbundantAtmosphericMethane, zahnle2020CreationEvolutionImpactgenerated, huang2022AssessmentAmmoniaBiosignature, woitke2021CoexistenceCH4CO2}, and again here we see that the behaviour of these gases is central to understanding the chemistry of volcanic atmospheres on hot rocky exoplanets. The production of \ce{NH3} and \ce{CH4} from high-temperature volcanic precursors is ultimately dependent on both the chemical timescales of the reactions, and the time period over which gases have to relax (either the time period of observation, or the cooling rate of the gas if the temperature is changing). Our next step is therefore to constrain the kinetics of these reactions, in order to understand how high temperature volcanic gas chemistry translates to lower temperature planetary atmospheres.

\subsection{Timescales to thermochemical equilibrium}
\label{section:kinetics}

\begin{figure}[ht]
    \centering
    \includegraphics[width=0.5\textwidth]{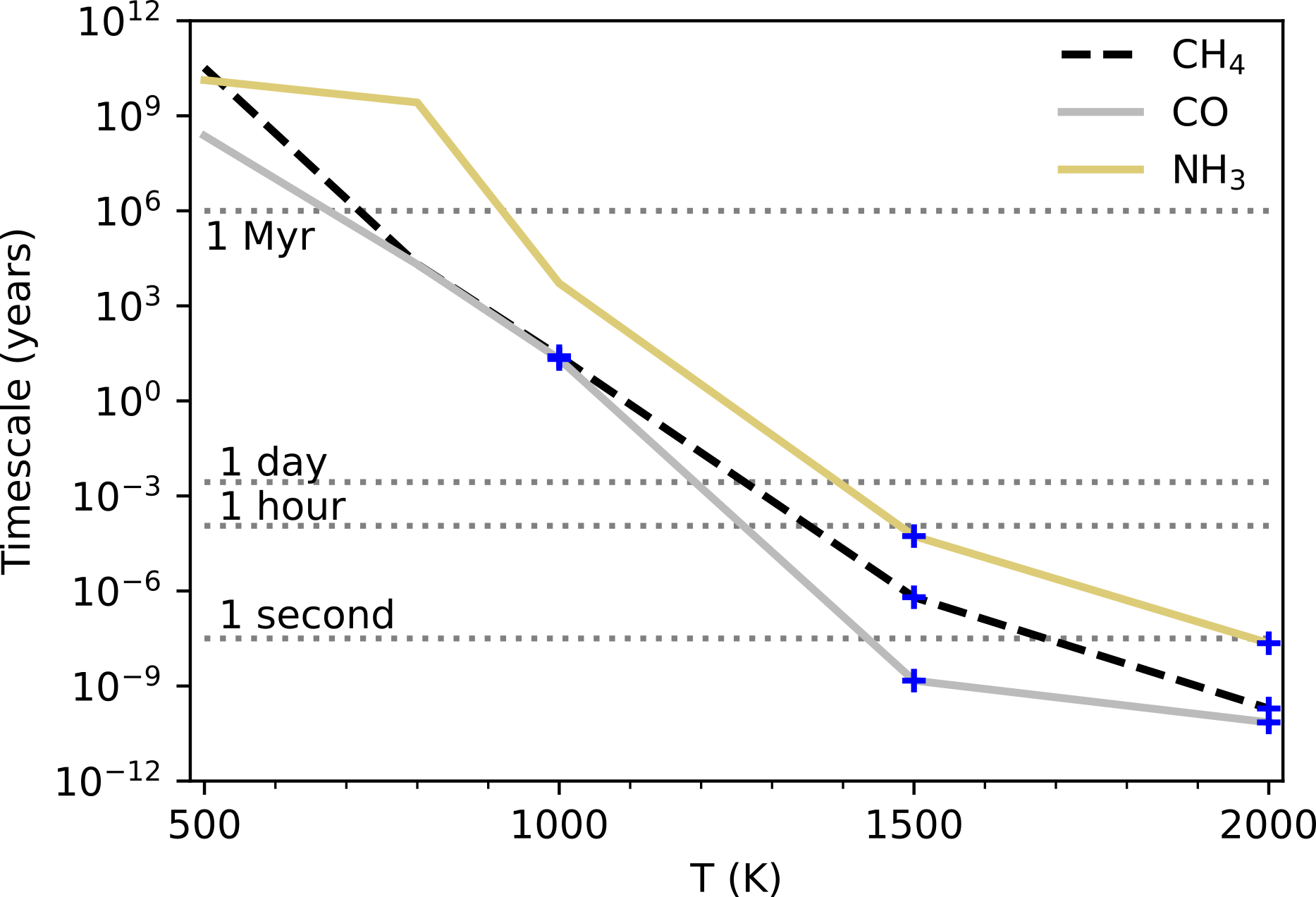}
    \caption{The chemical timescales for \ce{CH4}, CO and \ce{NH3} to reach thermochemical equilibrium for instantaneous cooling/heating of the 0D atmosphere at 23\,bar from a starting temperature of 1473\,K. Timescales were evaluated for final temperatures of 500, 800, 1000, 1500 and 2000\,K. The starting atmospheric composition is that produced by 1\,Gyr of degassing from a mantle with an $f\mathrm{O}_2$ = IW. Where thermochemical equilibrium was reached by ARGO (indicated by blue crosses) the time taken to reach equilibrium is plotted; otherwise, instantaneous chemical timescales are used to estimate time to equilibrium. Timescales are given in years, and are estimated to vary over nearly 24 orders of magnitude.}
    \label{fig:kinetics}
\end{figure}

The chemical timescales to new equilibrium abundances for \ce{CH4}, CO and \ce{NH3} are plotted in Fig.\,\ref{fig:kinetics}. These calculations show an atmospheric composition produced by volcanic degassing from a mantle with $f\mathrm{O}_2$ equal to the iron-w{\"u}stite (IW) buffer. This atmosphere has then been instantaneously cooled/heated to the given temperature from its magmatic temperature of 1473 K. We note that a physical scenario where the atmospheric temperature is significantly hotter than that of magmatic gases (and therefore the interior of the planet) is unlikely, but it is nonetheless useful to test the reaction kinetics at higher temperatures.

At temperatures of 1100\,K and higher, the atmosphere (at surface pressure, 23\,bar in this case) will re-equilibrate in less than one hundred years, a geological blink of an eye. Above 1500\,K, these timescales drop to minutes, down to less than a second at 2000\,K. With timescales this short, processes other than chemical kinetics will be the rate limiting step for thermochemical equilibrium being reached within the atmosphere, such as the mixing/circulation time for the atmosphere. At 800\,K, while \ce{CH4} and CO re-equilibriate on the order of thousands of years, \ce{NH3} will now take 2.6\,Gyr to reach its new equilibrium abundance. However, once the surface temperature has dropped to 500\,K, all three species take on the order of a Gyr to reach their equilibrium abundances. At this temperature, still well above what would usually be considered `habitable', \ce{CH4} has a chemical timescale of 33\,Gyr to equilibrium; over 30$\times$ longer than the period of volcanic outgassing which produced the atmosphere. At 500\,K and without reaction catalysis, this atmosphere would therefore take more than double the age of the universe to fully re-equilibrate.

\begin{figure}[ht]
    \centering
    \includegraphics[width=\textwidth]{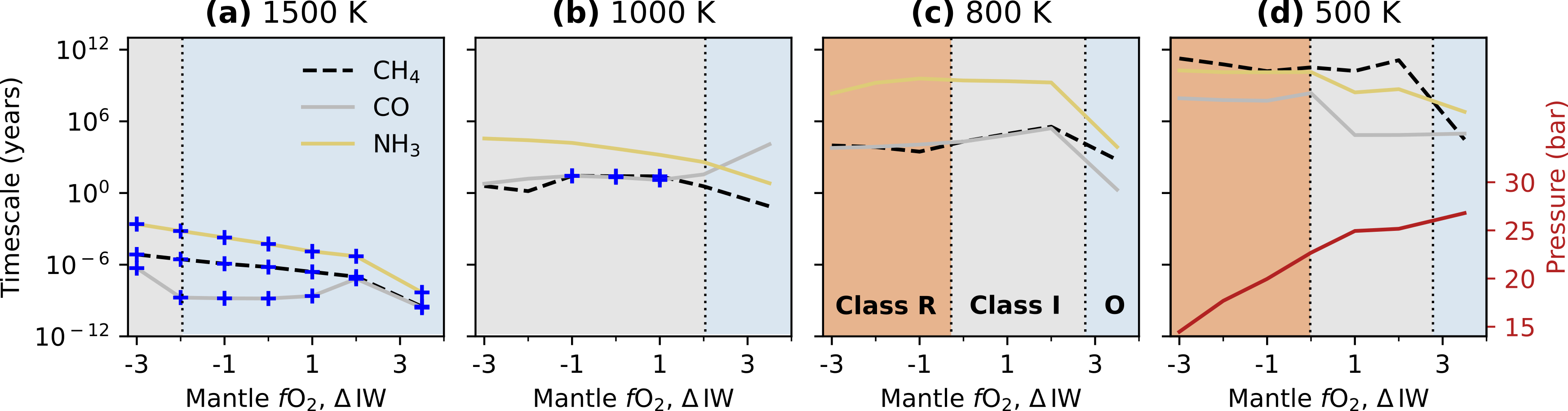}
    \caption{Estimates of the chemical timescales for \ce{CH4}, CO and \ce{NH3} to reach thermochemical equilibrium for instantaneous cooling/heating of the 0D atmosphere from 1473\,K to 1500, 1000, 800 and 500\,K respectively. Blue crosses indicate simulations where ARGO achieves the FastChem equilibrium number density of the given species. Timescales are given in years, increasing from minutes at 1500\,K, to hundreds of Gyr at 500\,K. Background colours denote the atmospheric classes derived in Paper I \cite{liggins2022GrowthEvolutionSecondary} for the fully re-equilibrated atmosphere at each temperature, labeled in (c). The surface pressure (P$_\mathrm{surf}$) trend (identical for all 4 temperature scenarios) is plotted as a red line in (d).}
    \label{fig:fo2_dependence}
\end{figure}

The speciation of volcanic gases, and therefore the atmospheres they form, is dependent on the mantle $f\mathrm{O}_2$ of the planet (see Fig.\,\ref{fig:fo2_dependence} and Paper I for more discussion). Fig.\,\ref{fig:fo2_dependence} shows how the timescales to equilibrium shown in Fig.\,\ref{fig:kinetics} are affected by the different atmospheric conditions imposed by varying the $f\mathrm{O}_2$ of the mantle from which outgassing has occurred, at 1500, 1000, 800 and 500\,K.

The different atmospheric speciations and surface pressures that result from varying mantle $f\mathrm{O}_2$ have minimal effects on the results already described (although this ignores any climate effects induced by the changing atmospheric speciation, see Sect.\,\ref{section:caveats}). Over 9 log-units of $f\mathrm{O}_2$ change, the timescales to equilibrium stay broadly grouped at similar values; at 1500\,K, full thermochemical equilibrium is reached after a maximum of around 1 week at IW-3, down to a few minutes at IW+3.5 (and less than a second at IW+2). In contrast, at 500\,K the very fastest timescales are around 0.2 Myr, while at lower mantle $f\mathrm{O}_2$ conditions \ce{NH3} and \ce{CH4} have chemical timescales of 10's to 100's of Gyr. It should be noted that these values are estimated timescales, which, as discussed in Section \ref{section:methods}, may overestimate the time to equilibrium if the solver has stalled close to the true equilibrium composition. However, reducing the timescales of most of the $f\mathrm{O}_2$ range shown in Fig.\,\ref{fig:fo2_dependence}d even by a factor of 10-100 would still result in an atmosphere at disequilibrium after several Gyr: i.e., comparable to the age of planets we will observe over the coming decades.

Aside from the trend of timescale with temperature, there is also a slight trend to shorter timescales under more oxidised conditions (Fig.\,\ref{fig:fo2_dependence}), particularly at lower temperature conditions. This is a result of the higher pressure volcanic atmospheres generated by outgassing of more oxidised mantles (see Paper I and Fig.\,\ref{fig:fo2_dependence}d) and hence more frequent molecular collisions. Linking these results back to the atmospheric classes we introduced in Paper I for an 800\,K atmosphere, Class O (oxidised mantle) atmospheres tend to have faster, but more variable rates to equilibrium at a given surface temperature, while Classes I and R (intermediate and reduced mantles, respectively) have slightly slower, but more consistent timescales.

\subsection{Rate-limiting Reactions}
\label{section:rates}

The key abundant and redox sensitive species exhibiting quenching are CO, \ce{CH4} and \ce{NH3}. The interconversions of \ce{CO <=> CH4} and \ce{N2 <=> NH3} proceed via networks of reactions with multiple intermediate steps and several possible reaction pathways. In this web of reactions, the kinetically limiting step converting one species to another is that step which progresses at the slowest rate along the overall fastest reaction pathway. The model discussed here ignores any inhibiting or catalysing processes such as photochemistry or atmosphere-surface reactions; accordingly, identifying the limiting step in each interconversion can indicate if, and via which step directly, such processes might affect the rate of re-equilibration.

The dominant reaction pathways and limiting reactions for both the \ce{CO <=> CH4} and \ce{N2 <=> NH3} interconversions have been debated in the literature, with the primary focus being on explaining disequilibrium chemistry observed in the atmospheres of giant planets (e.g., HD189733b, HD209458b, GJ436b, Gliese 229b) \cite{Moses2011,VisscherMoses2011,Line2011,ZahnleMarley2014}. \citeA{Tsai2017,Tsai2021} perform a detailed model intercomparison for the above conversions and demonstrate that the differences between the results of previous studies are primarily due to different choices of reaction rate coefficients for key reactions, despite the use of a less extensive chemical network by \citeA{Tsai2017,Tsai2021}. The rates of each reaction involved in the overall conversion of \ce{CO} and \ce{N2} depend on often poorly constrained rate coefficients; a further motivation for performing a detailed analysis into the reaction pathways and rate-limiting steps is to identify the underlying experimentally/theoretically constrained rate coefficients that drive our results. A more detailed description and analysis of the reactions and rates we discuss, including full reaction pathways and temperature and $f\mathrm{O}_2$ dependencies, can be found in \ref{section:c_appendix} and \ref{section:n_appendix} for the \ce{CO -> CH4} system and \ce{N2 -> NH3} system, respectively. In the following two sections we now briefly examine the reaction pathways and rate-limiting steps that are ultimately responsible for the quenching of \ce{CO}, \ce{CH4}, and \ce{NH3} in our reaction network.

\subsubsection{\ce{CO <=> CH4}}

\begin{figure}[ht]
    \centering
    \includegraphics[width=\textwidth]{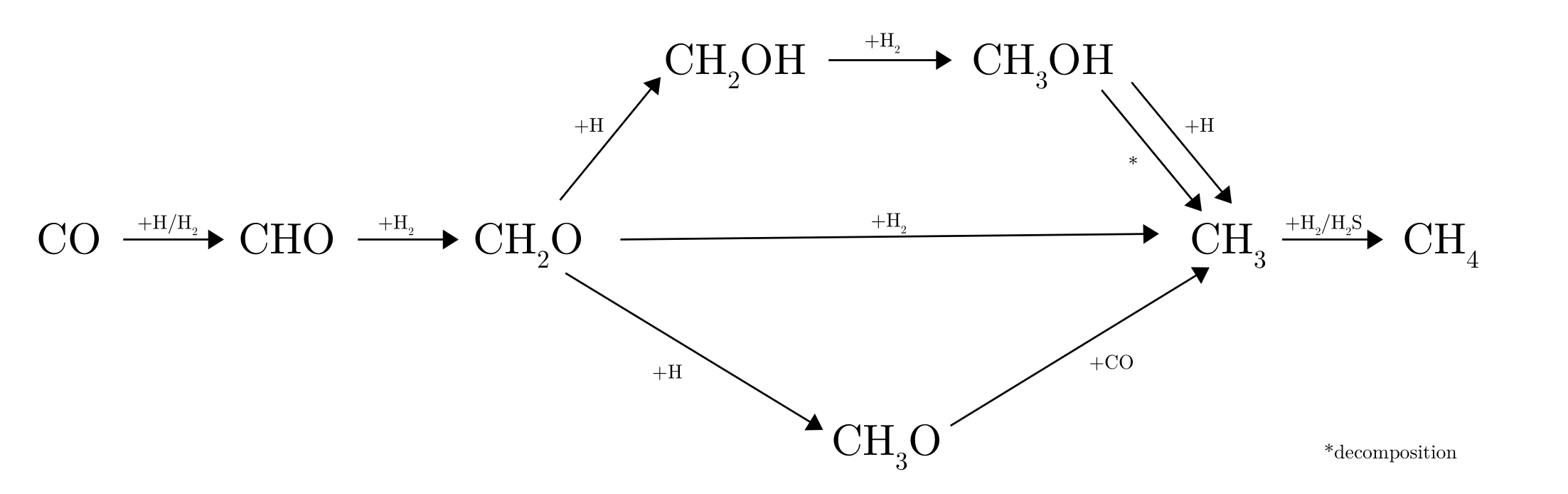}
    \caption{Network of important reactions linking \ce{CO} to \ce{CH4} in our chemical-kinetics network.}
    \label{fig:CO_CH4_network_main}
\end{figure}

In Fig.\,\ref{fig:kinetics} we showed that \ce{CO} and \ce{CH4} do not reach their equilibrium abundances in under 1\,Myr at $700\,-\,800\,{\rm K}$ and below. This is due to the slowest step along the reaction pathway causing a bottleneck in the conversion at these temperatures. The major chemical pathways linking \ce{CO} to \ce{CH4} are shown in Fig.\,\ref{fig:CO_CH4_network_main}. We find that the dominant reaction pathway for the \ce{CO -> CH4} conversion changes as a function of temperature but maintains the same net reaction:
\begin{linenomath*}
\begin{align}
	\ce{CO + 3H2 &-> CH4 + H2O}. \label{eqM:cycle_CO-CH4_1}
\end{align}
\end{linenomath*}
At 1000\,K and 800\,K, we find that the rate-limiting step of the dominant reaction pathway for the \ce{CO -> CH4} conversion (rate coefficient, $k$ (cm$^3$\,s$^{-1}$), taken from \citeA{rimmerCHEMICALKINETICSNETWORK2016}) is
\begin{linenomath*}
\begin{align}
    \ce{CH3OH + H -> CH3 + H2O} \label{eqM:Our_CH4_limiting}\\
    k = 9.41\times10^{-9}\exp(-124000/{\rm T}),
\end{align}
\end{linenomath*}
corresponding to the breaking of the \ce{C-O} bond. This set of reactions has also been identified as important for giant planets \cite{Yung1988, Visscher2010, Line2011}.  While others have identified alternative rate-limiting steps, the difference is likely due to the inclusion of photochemistry in their systems \cite{Yung1988, Visscher2010}, as e.g., \citeA{tsai2018} identify the same rate-limiting step as eq.\,\ref{eqM:Our_CH4_limiting} for equivalent temperature and pressure conditions.

At 500\,K, we find that the rate-limiting step of the dominant reaction pathway for the \ce{CO -> CH4} conversion is
\begin{linenomath*}
\begin{align}
    \ce{CH3O + CO -> CH3 + CO2} \label{eqM:Our_secondary_CH4_limiting}\\
    k = 2.61\times10^{-11}\exp(-5940/{\rm T}),
\end{align}
\end{linenomath*}
with the rate coefficient taken from \citeA{TsangHampson1986}. This pathway has not been discussed in previous literature regarding the \ce{CO <=> CH4} interconversion. 

At $800\,{\rm K}$, in the most reduced case, $f\mathrm{O}_2$\,=\,IW-3, the pathway in eq.\,\eqref{eqM:Our_secondary_CH4_limiting} is one order of magnitude slower than reaction \eqref{eqM:Our_CH4_limiting}. However, this reaction increases in importance for volcanic atmospheres built from increasingly oxidised mantles, and becomes comparable in rate to reaction (\ref{eqM:Our_CH4_limiting}) by $f\mathrm{O}_2$\,$\geq$\,IW. At $500\,{\rm K}$, the reaction rates become extremely small and the pathway along  \eqref{eqM:Our_secondary_CH4_limiting} becomes almost entirely responsible for the little production of \ce{CH4} that occurs. This is the reaction ultimately responsible for the quenching of CO and \ce{CH4} abundances at low temperatures. The relative importance of all the significant reaction pathways as a function of atmospheric temperature and mantle $f\mathrm{O}_2$ can be found in Fig.\,\ref{fig:ch4_pathways}.

\subsubsection{\ce{N2 <=> NH3}}

\begin{figure}[ht]
    \centering
    \includegraphics[width=\textwidth]{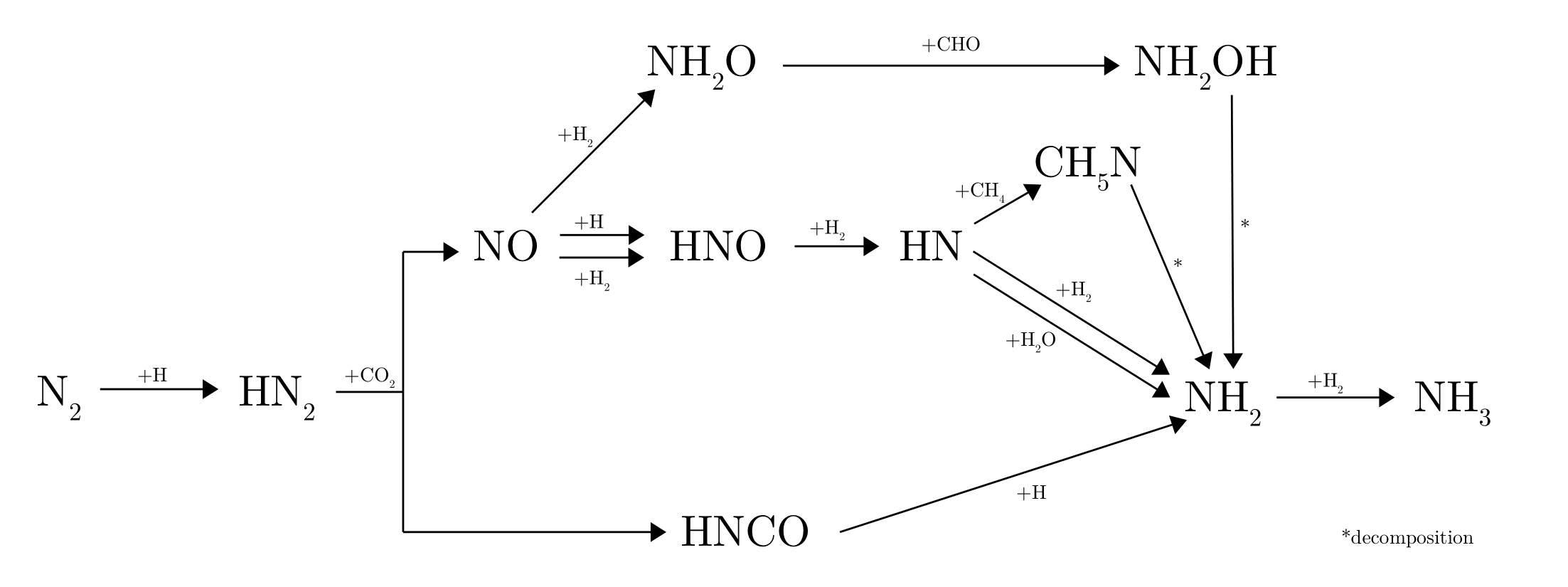}
    \caption{Network of important reactions linking \ce{N2} to \ce{NH3} in our chemical-kinetics network.}
    \label{fig:N2_NH3_network_main}
\end{figure}

The interconversion of \ce{N2 <=> NH3} has historically been more difficult to assess compared to that of \ce{CO <=> CH4}, with less well understood reaction schemes \cite{Moses2010,Moses2011}. We have found that at temperatures $\leq$ 1000\,K, \ce{NH3} does not reach its equilibrium abundance after 1\,Myr, due to the slowest step in the interconversion causing a bottleneck at these temperatures (Fig.\,\ref{fig:kinetics}). The major chemical pathways linking \ce{N2} to \ce{NH3} are shown in Fig.\,\ref{fig:N2_NH3_network_main}. 

We find that the dominant reaction pathway for the \ce{N2 -> NH3} conversion remains the same over all temperatures, with net reaction:
\begin{linenomath*}
\begin{align}
	\ce{N2 + 3H2 &-> 2NH3}, \label{eqM:cycle_N2-NH3_combined}
\end{align}
\end{linenomath*}
and the rate-limiting step and coefficient \cite{Tomeczek2003} of
\begin{linenomath*}
\begin{align}
    \ce{HN2 + CO2 -> HNCO + NO} \label{eqM:Our_NH3_limiting},\\
     k = 3.32\times10^{-11}\rm{exp}(-12630\,K/T)\label{eq:Our_NH3_limiting}.
\end{align}
\end{linenomath*}
Reaction \ref{eqM:Our_NH3_limiting} is endothermic and therefore the numerator of the exponent in eq.\,\eqref{eq:Our_NH3_limiting} depends on the change in Gibbs free energy, which changes as a function of temperature. Here we evaluate it at $\sim\,900\,{\rm K}$, the approximate quench temperature found for \ce{NH3}.  We note the rate constant we use for reaction \mbox{\eqref{eq:Our_NH3_limiting}} assumes no barrier, therefore if there is a significant activation barrier to \mbox{$\ce{HN_2}+\ce{CO2} -> \ce{HNCO} + \ce{NO}$}, then another reaction may set the rate limiting step or overall conversion pathway.  However, we note (1) that even if when we remove this reaction from the network altogether, we obtain the nearly the same timescale, and (2) consistent with this observation is that similar results have been found by different models with different chemical networks.  Together, these imply that \mbox{\ce{N2}--\ce{NH3}} quenching is insensitive to the precise details of the reaction scheme.

While the net reaction \eqref{eqM:cycle_N2-NH3_combined} matches that of previous schemes \cite{Line2011,Moses2011,tsai2018}, the mechanism of conversion differs. Instead of sequential reactions of \ce{N2} with \ce{H} and \ce{H2} \cite{Moses2011,Line2011}, our reaction scheme follows the addition of a \ce{H} atom with the breaking of the \ce{N\bond{=}N} bond, resulting in two separate branches, one following \ce{HCNO}, and the other following \ce{NO} (See Fig.\,\ref{fig:N2_NH3_network_main} and \ref{section:n_appendix}).

At 2000\,K and 1500\,K both branches contribute, and the pathways are limited by reaction \eqref{eqM:Our_NH3_limiting}. As the temperature lowers, further limiting reactions emerge along the NO branch and this becomes subdominant compared to the HCNO branch, which remains limited by \eqref{eqM:Our_NH3_limiting} over our whole temperature and $f\mathrm{O}_2$ range. Reaction \eqref{eqM:Our_NH3_limiting} is thus the reaction ultimately responsible for the quenching of \ce{NH3} abundances at low temperatures. The relative importance of all the significant reaction pathways as a function of atmospheric temperature and mantle $f\mathrm{O}_2$ can be found in Fig.\,\ref{fig:nh3_pathways}.

\section{Discussion}
\label{section:discuss}

Our results demonstrate that for a volcanically-derived atmosphere, the surface temperature has to be in excess of 1000\,K in order for the chemical timescales of all three species considered here to be shorter than 1\,Myr, across all of the mantle $f\mathrm{O}_2$-dependent atmospheric classes identified in Paper I. 
As this is 1000\,$\times$ shorter than the age of the atmospheres considered in Figs.\,\ref{fig:kinetics} \& \ref{fig:fo2_dependence} (which are calculated for an atmosphere 1\,Gyr old), we consider chemical timescales of 1\,Myr to be the cutoff point for assuming equilibrium, where longer timescales indicate that the abundance of the species in the atmosphere will start to significantly diverge from the abundance predicted by thermochemical equilibrium.

When this assumption is applied to the results shown in Figs.\,\ref{fig:kinetics} \& \ref{fig:fo2_dependence}, it can be seen that there is a temperature range where only \ce{NH3} has a calculated timescale greater than 1\,Myr, while CO and \ce{CH4} have timescales shorter than this cutoff. This reflects a limiting-step reaction rate for the \ce{N2 <=> NH3} conversion which is systematically many orders of magnitude lower than that of the \ce{CO <=> CH4} conversion.
In this scenario, the atmosphere can be thought of as being in partial equilibrium; where C-H-O species, forming the bulk of the atmosphere, should be approaching thermochemical equilibrium, but major nitrogen species such as \ce{NH3} have their abundances dictated by their freezeout temperatures, reflecting a higher temperature equilibrium.

At higher temperatures, $\geq$\,1000\,K, all species have timescales shorter than 1\,Myr and so the atmosphere is assumed to be closely approaching thermochemical equilibrium (discounting photochemistry and vertical mixing). At lower temperatures, below $\sim$\,700\,K, \ce{NH3}, CO and \ce{CH4} all have timescales longer than 1\,Myr so we assume an entirely `kinetically limited’ volcanic atmosphere where even over very long timescales, the atmosphere will no longer reflect thermochemical equilibrium, and instead the kinetics of different reactions control the atmospheric speciation.

\begin{figure}[h]
    \centering
    \includegraphics[width=0.7\textwidth]{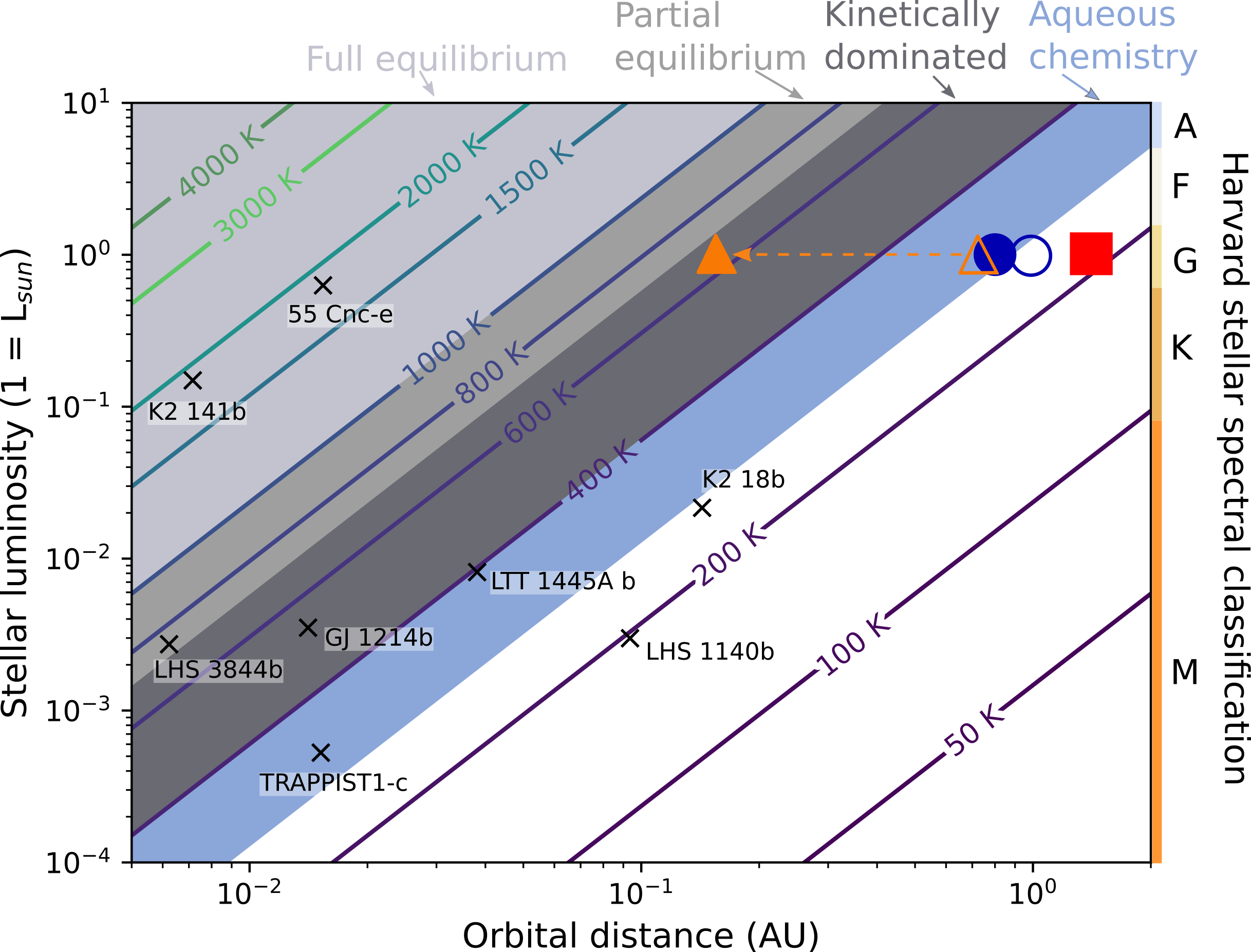}
    \caption{Regions in stellar luminosity--orbital distance space where volcanic gases injected into atmospheres may reach either full, partial or minimal thermochemical equilibrium at a given atmospheric temperature. Temperatures are plotted as the equilibrium temperature of a planet assuming a Bond albedo of 0.3. Atmospheres in partial equilibrium will have species such as \ce{NH3} with abundances dictated by their `freezeout' temperatures, while the rest of the species will be present at abundances predicted by thermochemical equilibrium for that temperature. The liquid water zone is also marked, indicating the point where aqueous chemistry will become relevant for influencing atmospheric chemistry and determining the lifetime of atmospheric species. Earth (blue circle) and Venus (orange triangle) and Mars (red square) are plotted both at their correct orbital distances (open symbols), and their actual surface temperatures (filled symbols), the latter as controlled by the climatic conditions created by their atmospheres. Also marked are a number of rocky exoplanets (all due to be studied during the first round of JWST observation), plotted according to their orbital distance and stellar luminosity; exoplanet data taken from \url{www.exoplanet.eu}.}
    \label{fig:stellar_groups}
\end{figure}

The presence of different temperature zones where atmospheres may be in full/ partial/ kinetically limited thermochemical equilibrium requires that certain stars, and orbital zones around those stars, will be optimal for finding planets with atmospheres which are in specific equilibrium regimes. 
These zones are shown in Fig.\,\ref{fig:stellar_groups}, according to the equilibrium temperature of a planet relative to the luminosity of its star and its orbital distance. Also shown is the point at which water will start to condense from the atmosphere, allowing aqueous chemistry within water droplets. Figure\,\ref{fig:stellar_groups} indicates that many exoplanets orbiting M stars (which are the most amenable to observations of secondary atmospheres, due to the low planet--star size ratio) will have to have a significant greenhouse effect to have atmospheres which can be modelled as approaching thermochemical equilibrium near their surfaces.

The cutoff temperatures presented in Fig.\,\ref{fig:stellar_groups} are based on timescales for the end-member case of instantaneously cooling a 0D atmosphere from its high temperature state of thermochemical equilibrium. However in nature, atmospheres will progressively gain mass over time, unless they reach a steady state where gas input is matched by removal processes (e.g., hydrogen escape, see Paper III). At any given time an atmosphere is thus a mix of older gases which have had time to approach equilibrium at surface temperatures, along with newly erupted gases which still reflect their high temperature origins. As volcanic gases tend to have characteristics of more oxidised atmospheres at higher temperatures, according to the class definitions of Paper I \cite{liggins2022GrowthEvolutionSecondary}, quenching the chemistry of hot volcanic gases and mixing them into the atmosphere will make the mantle $f\mathrm{O}_2$ appear more oxidised if the volcanic atmosphere was interpreted as being in thermochemical equilibrium (e.g., Fig.\,\ref{fig:fo2_dependence}, also see Figure B1 of Paper I).

Even on planets sitting within the `full equilibrium' regime in Fig.\,\ref{fig:stellar_groups}, rapid alteration processes such as escape (see Paper III) and photochemistry will be occurring in the atmosphere, which are not accounted for here. Photochemical processing will be particularly important on planets which are subject to intense UV radiation but have atmospheres in the partial equilibrium zone, where the surface temperatures are high enough to approximate equilibrium. In these cases the stratosphere may still be cool enough that photochemistry can significantly affect the abundances of observable species. Surface-atmosphere interactions and any potential biological activity have also been neglected in this study, along with any possible catalysis of reactions. Each of these processes will likely act to pull the chemistry of an atmosphere away from thermochemical equilibrium, and towards some new steady-state. Given that, the results presented here likely represent the most favourable case for volcanic atmospheres achieving thermochemical equilibrium. The suggested temperature zones for partial equilibrium and kinetic dominance (shown in Fig.\,\ref{fig:stellar_groups}) should therefore be taken as the coolest possible temperature estimations of such boundaries.

\subsection{Can volcanic atmospheres produce biosignature false positives?}

Results presented here suggest that 700\,K is the minimum atmospheric temperature where a volcanic atmosphere might in principle be able to relax to thermochemical equilibrium (if given $\leq1$\,Myr to do so). Above this temperature, it is appropriate for the atmosphere to be modelled as being close to thermochemical equilibrium. This limit is presented for the simplest, most favourable case for achieving equilibrium, as the surface pressure (the highest pressure found in an atmosphere) is assumed for the reaction conditions, and all factors confounding attainment of local thermochemical equilibrium (e.g., transport, photochemistry) have been neglected. For atmospheres with surface temperatures much lower than 700\,K, such as some of those presented by \cite{herbort2020AtmospheresRockyExoplanets} and all in \cite{woitke2021CoexistenceCH4CO2}, significant amounts of catalysis must be assumed for the atmosphere to remain in thermochemical equilibrium. The results of this paper therefore bring into question the validity of using purely thermochemical modelling when considering planets with cooler surface temperatures.

Our results also support the use of co-occurring \ce{CH4} and \ce{CO2} in the absence of CO as a biosignature \cite{krissansen-totton2018DisequilibriumBiosignaturesEarth}. \ce{CH4} and \ce{CO2} \emph{are} present together in our modelled volcanically-derived atmospheres \cite<mantle $f\mathrm{O}_2$ $<$\,IW+3, corresponding to Class I and R, see Paper I;>[]{liggins2022GrowthEvolutionSecondary}, but in these cases CO is also present at abundances between 10\,ppm and 0.1\%. To completely remove CO to below ppm abundances in the atmospheres modelled here, while retaining both \ce{CH4} and \ce{CO2}, the surface temperature has to be below $\sim$\,500\,K. As shown in Figs.\,\ref{fig:kinetics} and \ref{fig:fo2_dependence}, the timescales required to achieve thermochemical equilibrium, and therefore remove CO, at temperatures $\leq$500\,K are sufficiently long (over a Gyr at 500\,K) that such an abiotic atmosphere is implausible. This result is also in agreement with that of \citeA{huang2022AssessmentAmmoniaBiosignature}, who suggest that ammonia can act as a good biosignature gas on warm to cool planets; while the production of \ce{NH3} from volcanic gases is thermodynamically favoured at low temperatures (Fig.\,\ref{fig:t_change}), the timescales to equilibrium through the conversion of \ce{N2} to \ce{NH3} at temperatures below 1000\,K are longer than the age of most planets. A volcanic source of ammonia in warm or cool atmospheres is therefore implausible.

In contrast, the claim of \cite{wogan2020AbundantAtmosphericMethane} that methane-rich atmospheres cannot be generated through volcanism may not hold for hot terrestrial atmospheres. Certainly, the results of this paper and Paper I \cite{liggins2022GrowthEvolutionSecondary} indicate that when considered over geologically relevant timescales and in atmospheres of between 1000 and 700\,K, significant \ce{CH4} fractions can be built up to a maximum of around 10\,\% where the mantle $f\mathrm{O}_2$ $\leq$\,IW+2.5. However, without additional modelling of photochemical processing within the example atmospheres shown here, the two results cannot be directly compared. Of course, this does not affect their claims around a strengthened case for a methane biosignature (which we agree with, as discussed above); 700\,K is not a temperature that would usually be considered `habitable'.

\subsection{Model Limitations}
\label{section:caveats}

There are a number of potentially important processes not included in our modelling that should be investigated in future work.

\textbf{Evolving atmospheric pressure}. The results presented here omit the effect of lower atmospheric pressures earlier in the planet’s lifetime, while the atmosphere is still growing. These lower pressures would promote correspondingly longer timescales to reach equilibrium, with the reaction rates then increasing with atmospheric growth and the associated increase in surface pressure.

\textbf{Self-consistent climate}. Our modelled volcanic atmospheres are not calculated with a self-consistent climate model, meaning that the temperature zones relative to stellar luminosity and orbital distance shown in Fig.\,\ref{fig:stellar_groups} are liable to move according to the speciation of the atmosphere. For example, \ce{CO2}-rich atmospheres such as those produced by oxidised and intermediate mantle $f\mathrm{O}_2$'s \cite{liggins2022GrowthEvolutionSecondary} will result in strong greenhouse warming, meaning planets with equilibrium temperatures presently plotted within the kinetically-dominated or aqueous zones of Fig.\,\ref{fig:stellar_groups} may be sufficiently greenhouse heated so as to have an atmospheric temperature suitable for equilibrium based modelling. Venus demonstrates this behaviour in Fig.\,\ref{fig:stellar_groups}, although with an atmosphere containing close to 90\,bar of \ce{CO2}, Venus is a somewhat extreme example. Climate effects would likely also result in increasing surface temperatures through time as a volcanically derived atmosphere grew.  The continued outgassing of greenhouse gases, such as \ce{CO2} and \ce{H2}, would mean a warming of the planet and an increasing reaction rate and progress towards thermochemical equilibrium over time.

\textbf{Dependence on chemical networks and chemical models}. The findings presented here can differ from previous studies that employ different chemical-kinetics networks for three primary reasons: inclusion of other disequilibrium processes not considered in this work (e.g., transport-induced quenching, photochemistry); the different redox ranges within the atmosphere (past studies often focus on the highly reducing hydrogen dominated atmospheres of irradiated giant planets, brown dwarfs, or post-impact worlds); and the use of different thermochemical data that determine reaction rate coefficients. It is possible that the inclusion of other disequilibrium processes may alter our results for rate-limiting reactions and quench temperatures of the \ce{CO <=> CH4} system and the \ce{N2 <=> NH3} system, however, disregarding photochemistry, the dominant pathways and rate-limiting reactions identified in Section\,\ref{section:rates} are mostly consistent with conclusions from recent previous studies \cite{zahnle2020CreationEvolutionImpactgenerated,ZahnleMarley2014,Moses2011}. Moreover, the quench temperatures identified for the \ce{CO <=> CH4} system and the \ce{N2 <=> NH3} system, $\sim 700\,{\rm K}$ and $\sim 1000\,{\rm K}$ respectively, are in agreement with results from \citeA{zahnle2020CreationEvolutionImpactgenerated} within $100\,{\rm K}$, who quote that quench temperatures are of the order $\sim 800\,{\rm K}$ and $\sim 1100\,{\rm K}$ respectively. The temperatures found by \citeA{zahnle2020CreationEvolutionImpactgenerated} relate to transport-induced quenching (when the vertical transport timescale by eddy diffusion becomes equal to the chemical timescale), whereas this study examines quenching purely due to the limit of chemical kinetics on planetary timescales ($\sim$ Myr). It is therefore expected that the quench temperatures we identify are systematically lower than the estimates of \citeA{zahnle2020CreationEvolutionImpactgenerated}: our results will apply generally as a lower limit on the quench temperature of any kinetically-limited system, within the accuracy of the thermochemical data used in the chemical-kinetics network.

We look specifically at the generality of our reaction timescale limits in the next section.

\subsubsection{Network-independent limits on reaction timescales}
\label{section:minimum_temps}

We can find a network-independent minimum estimate of reaction timescales, by noting that a necessary first step for transitioning from one equilibrium state to another is the thermal dissociation of a molecule. The atoms and radicals that form from this first dissociation may react with other molecules as the system transitions into a new equilibrium state, but the system will not change until that first thermal dissociation reaction takes place. Therefore, the most rapid thermal dissociation of a major species sets the minimum possible timescale for a system to transition into a new equilibrium state. As we saw above with our complete chemical network, it is intermediate reactions which ultimately provide the bottleneck on a conversion rate from, e.g., \ce{CO <=> CH4}, hence why considering only the first dissociation of a major species provides a lower bound on reaction timescales. To assess the extent to which our results are network-dependent, we provide an estimate for this first dissociation timescale.

The equilibrium species from Fig.\,\ref{fig:t_change} with the weakest bond strength (and which will therefore likely react first) is \ce{H2S}, with a S-H bond strength of $\approx 335$ kJ/mol \cite{rumble2022HandbookChemistryPhysics}. This approximates the activation energy for a collision with another molecule to break the S-H bond. The rate constant for this reaction (s$^{-1}$) is
\begin{linenomath*}
\begin{equation}
k = \langle n \sigma v \rangle e^{-E_a/RT},
\end{equation}
\end{linenomath*}
where $n$ (cm$^{-3}$) is the number density of the \ce{H_2S} molecule, $\sigma$ (cm$^2$) is the collisional cross-section, $v$ (cm/s) is the velocity of the molecules, $E_a$ (kJ/mol) is the activation energy, $T$ (K) is the gas temperature, and $R = 8.3145$ J mol$^{-1}$ K$^{-1}$ is the gas constant. 

\begin{figure}[!htb]
\centering
\includegraphics[width=0.5\textwidth]{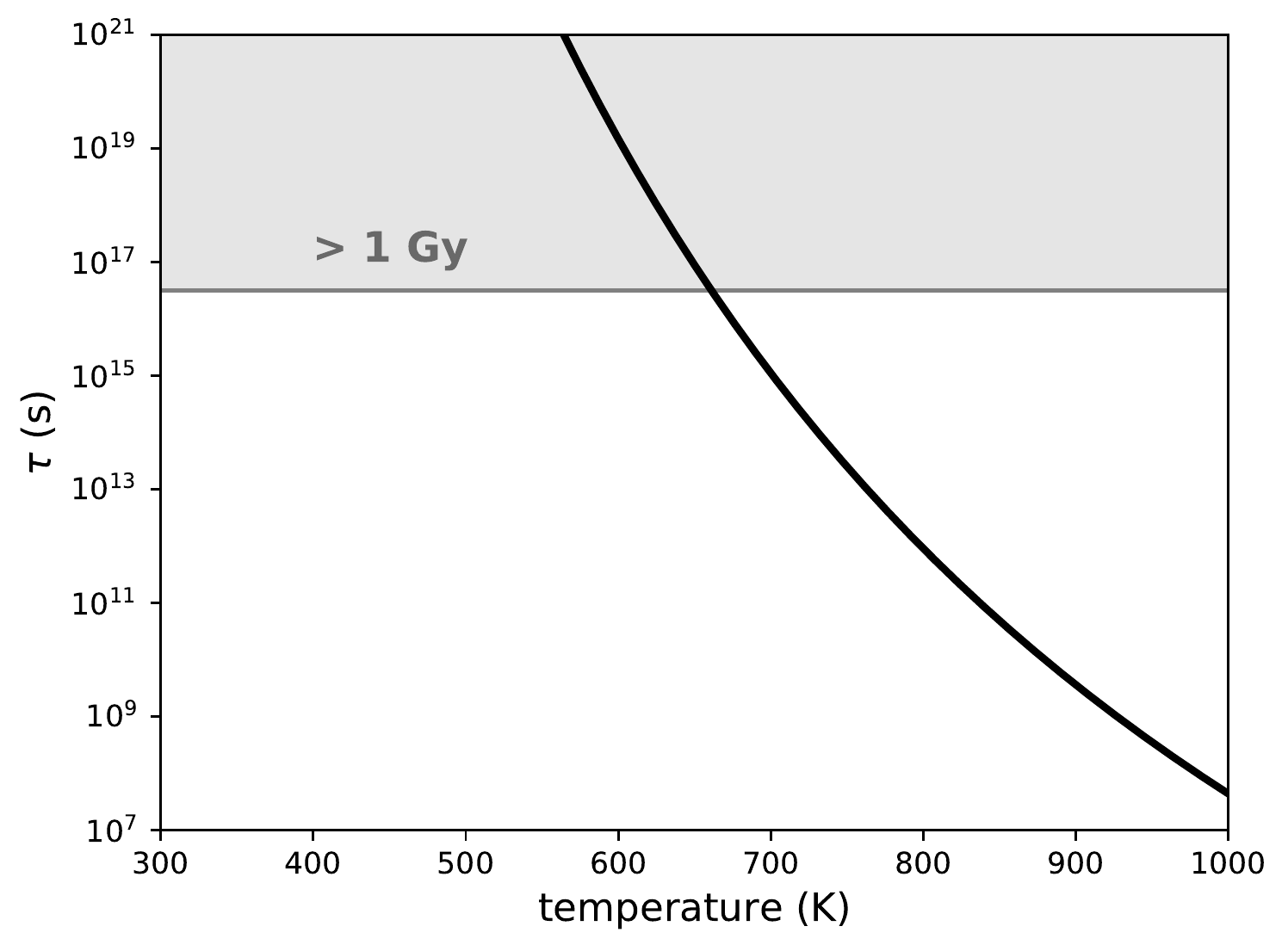}
\caption{Minimum chemical timescale $\tau$ (s) as a function of temperature (K). The black line is $\tau$, from Eq. (\ref{eq:tau-simple}), with $E_a = 335$ kJ/mol, based on the bond strength of S-H, the lowest bond strength out of the bonds for equilibrium species (see Fig.\,\ref{fig:t_change}). The grey line and shaded region show where timescales exceed 1\,Gy, with the crossover temperature at 661\,K. \label{fig:tau-simple}}
\end{figure}

We assume that the gas is in a Maxwell-Boltzmann distribution with a well-defined temperature. To provide an upper limit for the rate constant (and therefore a lower limit for the reaction timescale), we choose an average velocity for a small molecule (in this case, \ce{H_2}) and a collisional cross-section equal to the geometric cross-section for a large molecule ($\sigma = 3 \times 10^{-15}$ cm$^2$, similar to Benzene), and set $n$ equal to the number density of the gas at the planet's surface, assuming a 1 bar ideal gas. The abundance of \ce{H_2S} will always be less than this, and therefore the rate constant will be proportionately lower. Applying all these assumptions, and noting that the timescale $\tau$ (s) $ = 1/k$:
\begin{linenomath*}
\begin{equation}
\tau = \dfrac{\sqrt{\pi m_p kT}}{2\sigma p} \, e^{E_a/RT},
\label{eq:tau-simple}
\end{equation}
\end{linenomath*}
where $p = 1$\,bar is the gas pressure at the surface, $m_p = 1.67 \times 10^{-24}$\,g is the proton mass, and $k = 1.38 \times 10^{-16}$ erg/K is Boltzmann's constant. We plot $\tau$ as a function of $T$ in Figure \ref{fig:tau-simple}.

The temperature at which it will take longer than 1\,Gyr for the weakest bond in the system to break is 661\,K (Fig.\,\ref{fig:tau-simple}). This is less than 50\,K lower than our minimum temperature of 700\,K for thermochemical equilibrium to be achieved over geological timescales, as found by applying the full chemical network. We therefore find that even with an alternative chemical network, or by assuming equilibrium can be achieved with chemical timescales of over 1\,Myr, a significantly lower temperature for maintaining atmospheric thermochemical equilibrium is physically unfeasible.

\section{Conclusions}
\label{section:concs}

This paper has investigated the timescales over which volcanically--derived atmospheres can relax to a new thermochemical equilibrium, as dictated by the current atmospheric temperature and pressure. We find that the speciation of volcanic atmospheres will be quenched over geological time to those set by higher temperature thermochemical equilibria, if the atmospheric temperature is much below 700\,K. Below this threshold, slow chemical kinetics preclude the reactions of key species CO, \ce{CH4} and \ce{NH3}, preventing the atmosphere from reaching low-temperature thermochemical equilibrium. As such, the modelling of temperate exoplanet atmospheres where thermochemical equilibrium is assumed must also invoke significant catalysis of reactions in order for results to be plausible. Temperate planets with warm, or cool, quenched volcanic atmospheres will likely not show clear fingerprints of the mantle $f\mathrm{O}_2$, particularly if the mantle is highly reduced: the volcanic gases from reducing mantles are a closer match to those from more oxidised mantles when they are still hot, meaning that if such atmospheres were interpreted according to the classes derived in Paper I (reduced--intermediate--oxidising) they would be classed as being derived from more oxidising mantles than they really were (e.g., Fig.\,\ref{fig:fo2_dependence}, also see Figure B1 of Paper I, \citeA{liggins2022GrowthEvolutionSecondary}). Quenching of volcanically-derived atmospheres at the relatively high temperature of 700\,K also precludes the production of \ce{CO2}+\ce{CH4} without CO, meaning that false-positive biosignatures are unlikely to be achieved in volcanic atmosphere by thermochemistry alone.



%
%
%
%

%
%

%

%

\section{Open Research}
No new experimental data were generated for this work. The data used in producing 
the figures in this work can be found in \citeA{liggins2022dataset}. The EVolve model used to produce the initial atmospheric compositions is identical to the one used for Paper I \cite{liggins2022GrowthEvolutionSecondary}, and is available both at \citeA{liggins_philippa_2021_5645666}, and on a GitHub repository at \url{https://github.com/pipliggins/EVolve} where new versions will also be hosted. The FastChem 2.0 model of \citeA{stock2021} is also freely available at \url{https://github.com/exoclime/FastChem}, where updates are hosted. The methods underlying the STAND2020 chemical network used here are described in \citeA{rimmerCHEMICALKINETICSNETWORK2016} and \citeA{hobbs2021SulfurChemistryAtmospheres}.

\acknowledgments
The authors would like to thank the two anonymous referees for helpful and insightful comments that helped improve the quality of this manuscript. This work was funded by the Embiricos Trust Scholarship from Jesus College, Cambridge, and a grant from the Institute of Astronomy, Cambridge. S.J. thanks the UKRI Science and Technology Facilities Council (STFC) for the PhD studentship funding (grant reference ST/V50659X/1). For the purpose of open access, the authors have applied a Creative Commons Attribution (CC BY) licence to any Author Accepted Manuscript version arising. P.B.R thanks the Simons Foundation for support under SCOL awards 59963.

\appendix

\section{Atmospheric evolution}
\label{section:relax_appendix}

\begin{figure}[h!]
	\begin{center}
        \includegraphics[width=\textwidth]{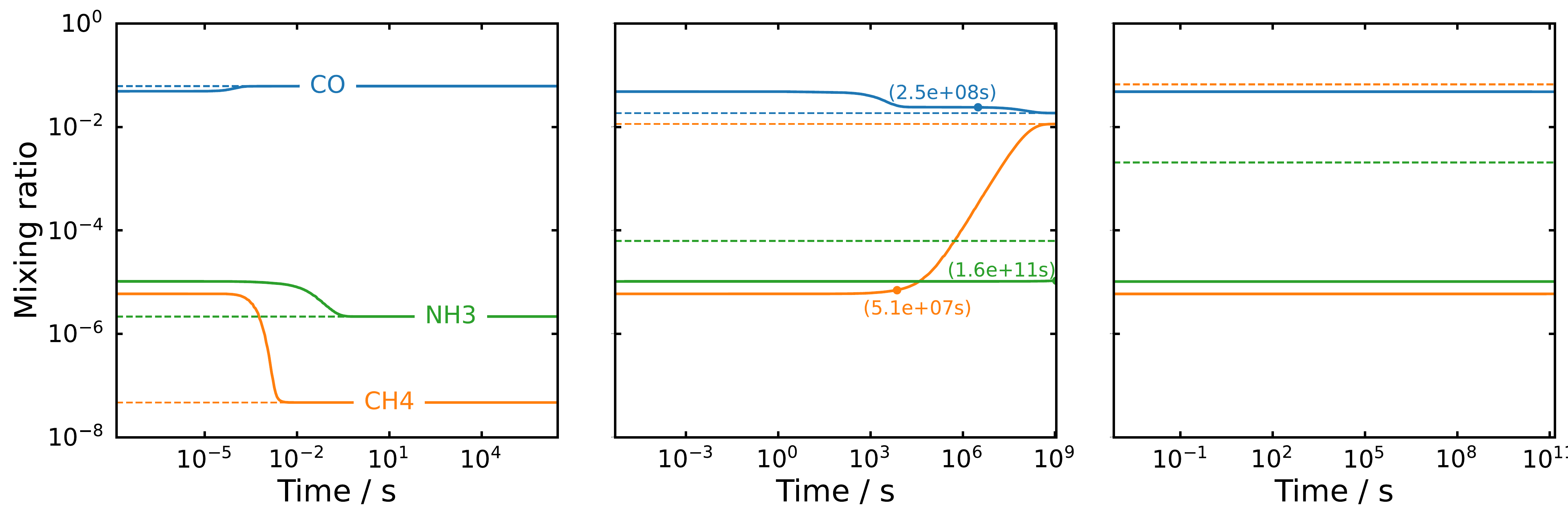}
	\caption{Time-series of the volcanic atmosphere composition produced on a planet with a mantle $f\mathrm{O}_2$ equal to IW as it relaxes to ambient atmospheric temperatures for three different cases: (left) hot (2000\,K), (middle) warm (1000\,K), and (right) cool (500\,K). The dashed lines show the equilibrium abundances of each species at the atmospheric temperature calculated by FastChem 2.0 \citeA{stock2021}. In the warm case we show, as numbers above filled circles along the lines, the estimated timescales to equilibrium that are computed before the mixing ratio approaches equilibrium.  The filled circles have been placed along the lines at the point where the species begins moving towards its final equilibrium abundance, or, in the case of \ce{NH3}, at the point where the calculation stalls.  The timescale estimates for \ce{CH4} and \ce{CO} provide an accurate order of magnitude estimate of the final time to equilibrium.}
	\label{fig:relax_hot_warm_cool}
\end{center}	
\end{figure}

In Fig.\,\ref{fig:relax_hot_warm_cool} we demonstrate how the atmospheric composition evolves for three atmospheric temperatures: (Fig.\,\ref{fig:relax_hot_warm_cool} left) hot (2000\,K), (Fig.\,\ref{fig:relax_hot_warm_cool} middle) warm (1000\,K), and (Fig.\,\ref{fig:relax_hot_warm_cool} right) cool (500\,K).

At 2000\,K, the volcanic gas composition relaxes to the atmospheric equilibrium composition on short timescales. In this `hot' case we can integrate the atmospheric chemistry at the new temperature until the 2000\,K equilibrium is reached.

At 1000\,K, the volcanic gas composition relaxes to the atmospheric equilibrium composition on intermediate timescales that are longer than in the hot atmosphere, but are shorter than geological timescales. For \ce{CO} and \ce{CH4} the time-series demonstrates that these species will reach equilibrium within the integration time, however for \ce{NH3} the timescales to equilibrium are longer than for \ce{CO} and \ce{CH4} and the solver stalls before \ce{NH3} reaches its true equilibrium. The timescale to the true equilibrium for \ce{NH3} must instead be calculated based on the rate of change of the \ce{NH3} number density and the difference between the final \ce{NH3} number density compared to its equilibrium number density at 1000\,K. This is calculated with equation \eqref{eq:timescale} at the latest time-step before stalling. In Fig.\,\ref{fig:relax_hot_warm_cool} we also show the timescale estimate that would be obtained for \ce{CO} and \ce{CH4} if the solver would have stalled at the quasi-steady state before they reached their true equilibrium. We find that the timescale estimates are consistent with the time that was actually taken for these two species to approach equilibrium.

At 500\,K, the volcanic gas composition relaxes to the atmospheric equilibrium composition on timescales that are longer than geological timescales. In this `cool' case the solver stalls at a quasi-steady state before any chemical species reaches its equilibrium. The time to the true equilibrium for \ce{CO}, \ce{CH4}, and \ce{NH3} is calculated to be longer than geological timescales (see section \ref{section:kinetics}). The species are thus said to be `quenched' at their hotter volcanic gas composition and will not reflect the atmospheric-temperature equilibrium composition unless there is significant catalysis due to other physical processes. In the absence of additional catalysis, the atmospheric composition of the cool secondary atmospheres will unintuitively resemble the equilibrium composition of much hotter atmospheres (i.e., at the magmatic temperature $\sim$ 1500\,K). The warm atmospheres, not the cool atmospheres, end up deviating the most from the hot equilibrium composition.

\section{Full reaction pathways and temperature dependence of the \ce{CO -> CH4} system}
\label{section:c_appendix}

In order to identify the dominant pathways and their rate-limiting steps, we use an algorithm that finds all significant routes from one molecule to another in our reaction network. This is done using the rates of reaction calculated in the chemical-kinetics simulation. Once all significant pathways are found, the rate-limiting step of each pathway is identified as the slowest reaction rate in the sequence of reactions. We do this for the pairs of start/end molecules \ce{CO -> CH4} and \ce{N2 -> NH3}, as a function of $f\mathrm{O}_2$ and temperature.

For the \ce{CO -> CH4} system, we identify four pathways that each dominate at different atmospheric temperatures.

Pathway A:
\begin{linenomath*}
\begin{align}
	\ce{CO + H + M &-> CHO + M},\\
	\ce{CHO + H2 &-> CH2O + H},\\
	\textbf{\ce{CH2O + H2}} &\textbf{\ce{-> CH3 + HO}}\label{eq:CH4_limiting_A},\\
	\ce{CH3 + H2 &-> CH4 + H},\\
	\ce{H + HO &-> H2O},\\
	{\rm Net:} \; \ce{CO + 3\,H2 &-> CH4 + H2O}.
\end{align}
\end{linenomath*}
Pathway A dominates the conversion at 2000\,K over all the $f\mathrm{O}_2$ values investigated. The pathway is  limited by reaction \eqref{eq:CH4_limiting_A}.

Pathway B:
\begin{linenomath*}
\begin{align}
	\ce{CO + H + M &-> CHO + M},\\
	\ce{CHO + H2 &-> CH2O + H},\\
	\ce{CH2O + H + M &-> CH2OH + M}\\
	\textbf{\ce{CH2OH + H2}} &\textbf{\ce{-> CH3OH + H}}\label{eq:CH4_limiting_B_2000K},\\
	\textbf{\ce{CH3OH}} &\textbf{\ce{-> CH3 + HO}}\label{eq:CH4_limiting_B},\\
	\ce{CH3 + H2S &-> CH4 + HS},\\
	\ce{H + HS &-> H2S},\\
	\ce{HO + H2 &-> H2O + H}.\\
	{\rm Net:} \; \ce{CO + 3\,H2 &-> CH4 + H2O}.
\end{align}
\end{linenomath*}
Pathway B dominates the conversion at 1500\,K over all the $f\mathrm{O}_2$ values investigated. The pathway is  limited by reaction \eqref{eq:CH4_limiting_B_2000K} at 2000\,K and by reaction \eqref{eq:CH4_limiting_B} at lower temperatures.

Pathway C:
\begin{linenomath*}
\begin{align}
	\ce{CO + H + M &-> CHO + M},\\
	\ce{CHO + H2 &-> CH2O + H},\\
	\ce{CH2O + H + M &-> CH2OH + M},\\
	\ce{CH2OH + H2 &-> CH3OH + H},\\
	\textbf{\ce{CH3OH + H}} &\textbf{\ce{-> CH3 + H2O}}\label{eq:CH4_limiting_C},\\
	\ce{CH3 + H2S &-> CH4 + HS},\\
	\ce{H + HS &-> H2S},\\
	\ce{H2 &-> H + H}.\\
	{\rm Net:} \; \ce{CO + 3\,H2 &-> CH4 + H2O}.
\end{align}
\end{linenomath*}
Pathway C dominates the conversion at 1000\,K and 800\,K over all the $f\mathrm{O}_2$ values investigated. The pathway is limited by reaction \eqref{eq:CH4_limiting_C}.

Pathway D:
\begin{linenomath*}
\begin{align}
	\ce{CO + H + M &-> CHO + M},\\
	\ce{CHO + H2 &-> CH2O + H},\\
	\ce{CH2O + H + M &-> CH3O + M},\\
	\textbf{\ce{CH3O + CO}} &\textbf{\ce{-> CH3 + CO2}}\label{eq:CH4_limiting_D},\\
	\ce{CH3 + H2S &-> CH4 + HS},\\
	\ce{H + HS &-> H2S},\\
	\ce{H2 &-> H + H},\\
	\ce{CO2 + H &-> CO + HO},\\
	\ce{HO + H2 &-> H2O + H}.\\
	{\rm Net:} \; \ce{CO + 3\,H2 &-> CH4 + H2O}.
\end{align}
\end{linenomath*}
Pathway D dominates the conversion at 500\,K over all the $f\mathrm{O}_2$ values investigated. The pathway is limited by reaction \eqref{eq:CH4_limiting_D}.

\begin{figure}[hb]
    \centering
    \includegraphics[width=0.85\linewidth]{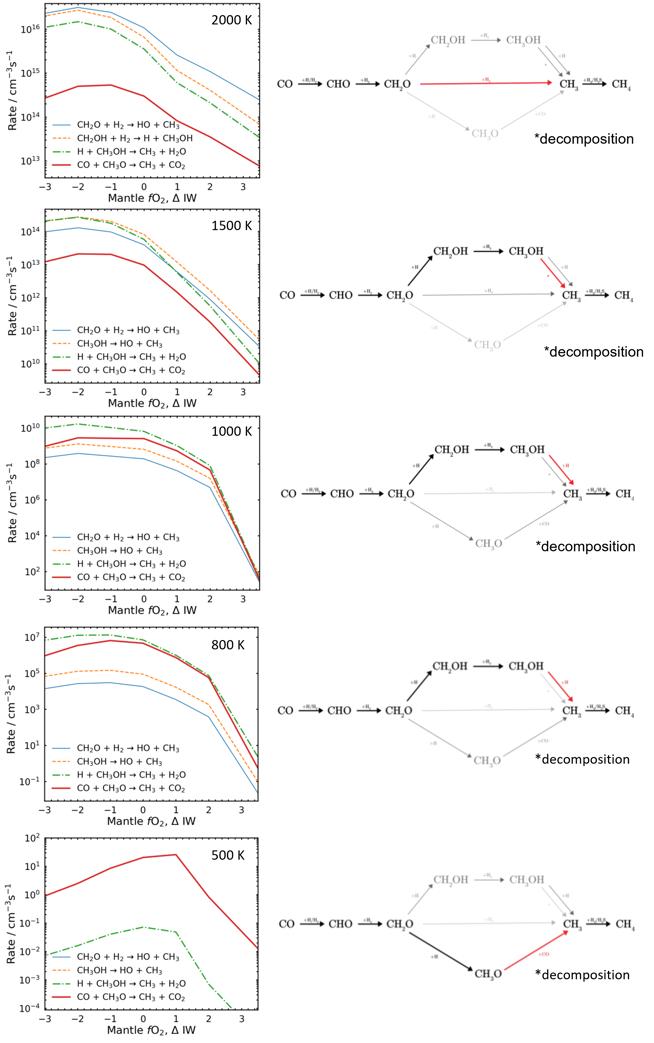}
    \caption{Reaction rates as a function of $f\mathrm{O}_2$ of the limiting reactions for pathways that dominate at one or more of the temperatures investigated. Alongside we show the dominant pathway highlighted in the reaction network in bold, and its limiting reaction highlighted in red. From top to bottom, the figures correspond to atmospheric temperatures of 2000\,K, 1500\,K, 1000\,K, 800\,K, and 500\,K.}
    \label{fig:ch4_pathways}
\end{figure}

The rates of the limiting reactions of each of the pathways A-D are shown in Fig.\,\ref{fig:ch4_pathways} as a function of mantle $f\mathrm{O}_2$ for atmospheric temperatures 2000\,K, 1500\,K, 1000\,K, 800\,K, and 500\,K. Where one pathway's limiting reaction is the fastest, that is the pathway that dominates the conversion. Annotated chemical pathways are shown highlighted in the reaction network alongside the reaction rate figures.

In each case the rate-determining step of the overall conversion is associated with breaking the \ce{C-O} single bond. This is in agreement with the conclusion of \citeA{Moses2011}, although they identify a different reaction as the rate-limiting step. \citeA{ZahnleMarley2014}, who perform a similar analysis to that of \citeA{Moses2011} for the atmospheres of self-luminous giants and brown dwarfs, find results that partially agree with the analysis of \citeA{Moses2011}, but only under the condition that \ce{CH4}$\gg$\ce{CO} in the atmosphere. The secondary rate limiting step found by \citeA{ZahnleMarley2014} matches our dominant rate-limiting reaction at 1000\,K and 800\,K (eq.\,\eqref{eqM:Our_CH4_limiting}). Since the volcanic atmospheres that we are focusing on in this study are not initially methane-rich, our rate-limiting step is consistent with the result of \citeA{ZahnleMarley2014}.

\clearpage

\section{Full reaction pathways and temperature dependence of the \ce{N2 -> NH3} system}
\label{section:n_appendix}

For the \ce{N2 -> NH3} system, we identify one pathway that dominates the conversion over all temperatures and $f\mathrm{O}_2$ values, another pathway (with three variants) that dominates alongside the first only at high temperatures, and two further pathways that are subdominant at different temperatures.

Pathway A:
\begin{linenomath*}
\begin{align}
	\ce{N2 + H + M &-> HN2 + M},\\
	\textbf{\ce{HN2 + CO2}} &\textbf{\ce{-> HNCO + NO}}\label{eq:NH3_limiting_A},\\
	\ce{HNCO + H &-> NH2 + CO},\\
	\ce{NH2 + H2 &-> NH3 + H}.\\
	{\rm Net:} \; \ce{N2 + CO2 + H2 + H &-> NH3 + NO + CO}.
\end{align}
\end{linenomath*}
Pathway A is  limited by reaction \eqref{eq:NH3_limiting_A}. Reaction \eqref{eq:NH3_limiting_A} produces two \ce{N}-containing products: \ce{HNCO} and \ce{NO}. This reaction splits the network into two branches. Pathway A follows the branch with \ce{HNCO}. Another pathway, B, with multiple variations depending on the intermediate reagents, follows the branch with \ce{NO}. At 2000\,K and 1500\,K both pathways A and B are limited by reaction \eqref{eq:NH3_limiting_A} and therefore both dominate the conversion at equal rates. At lower atmospheric temperatures, bottlenecks along the \ce{NO} branch emerge and the different pathways become secondary to pathway A. We show the variants of pathway B and further secondary reaction pathways below.

Pathway B$_1$:
\begin{linenomath*}
\begin{align}
	\ce{N2 + H + M &-> HN2 + M},\\
	\textbf{\ce{HN2 + CO2}} &\textbf{\ce{-> HNCO + NO}}\label{eq:NH3_limiting_B1_a},\\
	\ce{NO + H + M &-> HNO + M},\\
	\ce{HNO + H2 &-> HN + H2O},\\
	\textbf{\ce{HN + H2O}} &\textbf{\ce{-> NH2 + HO}}\label{eq:NH3_limiting_B1_b},\\
	\ce{NH2 + H2 &-> NH3 + H},\\
	\ce{HO + H2 &-> H2O + H}.\\
	{\rm Net:} \; \ce{N2 + CO2 + 3\,H2 &-> NH3 + HNCO + H2O}.
\end{align}
\end{linenomath*}
Pathway B$_1$ is limited by reaction \eqref{eq:NH3_limiting_B1_a} at temperatures of 1500\,K and hotter, and limited by reaction \eqref{eq:NH3_limiting_B1_b} at temperatures of 1000\,K and below.

Pathway B$_2$:
\begin{linenomath*}
\begin{align}
	\ce{N2 + H + M &-> HN2 + M},\\
	\textbf{\ce{HN2 + CO2}} &\textbf{\ce{-> HNCO + NO}}\label{eq:NH3_limiting_B2_a},\\
	\textbf{\ce{NO + H + M}} &\textbf{\ce{-> HNO + M}}\label{eq:NH3_limiting_B2_b},\\
	\ce{HNO + H2 &-> HN + H2O},\\
	\textbf{\ce{HN + H2}} &\textbf{\ce{-> NH2 + HO}}\label{eq:NH3_limiting_B2_c},\\
	\ce{NH2 + H2 &-> NH3 + H}.\\
	{\rm Net:} \; \ce{N2 + CO2 + 3\,H2 &-> NH3 + HNCO + H2O}.
\end{align}
\end{linenomath*}
Pathway B$_2$ is limited by reaction \eqref{eq:NH3_limiting_B2_a} at 2000\,K, and at 1500\,K for $f\mathrm{O}_2$=IW+2 and lower. The pathway is limited by reaction \eqref{eq:NH3_limiting_B2_c} at 1500K for $f\mathrm{O}_2$=IW+3.5. At 1000\,K the pathway is limited by reaction \eqref{eq:NH3_limiting_B2_b} for $f\mathrm{O}_2$=IW+1 and below, and limited by reaction \eqref{eq:NH3_limiting_B2_c} for $f\mathrm{O}_2$=IW+2 and above. At 800\,K the pathway is limited by \eqref{eq:NH3_limiting_B2_b} for $f\mathrm{O}_2$=IW and below, and limited by reaction \eqref{eq:NH3_limiting_B2_c} for $f\mathrm{O}_2$=IW+1 and above.

Pathway B$_3$:
\begin{linenomath*}
\begin{align}
	\ce{N2 + H + M &-> HN2 + M},\\
	\textbf{\ce{HN2 + CO2}} &\textbf{\ce{-> HNCO + NO}}\label{eq:NH3_limiting_B3_a},\\
	\textbf{\ce{NO + H2 + M}} &\textbf{\ce{-> HNO + M}}\label{eq:NH3_limiting_B3_b},\\
	\ce{HNO + H2 &-> HN + H2O},\\
	\ce{HN + H2 &-> NH2 + HO},\\
	\ce{NH2 + H2 &-> NH3 + H},\\
	\ce{H + H + M &-> H2 + M}.\\
	{\rm Net:} \; \ce{N2 + CO2 + 3\,H2 &-> NH3 + HNCO + H2O}.
\end{align}
\end{linenomath*}
Pathway B$_3$ is limited by reaction \eqref{eq:NH3_limiting_B3_a} at 2000\,K, and limited by reaction \eqref{eq:NH3_limiting_B3_b} at 1500\,K and below.

Pathway C:
\begin{linenomath*}
\begin{align}
	\ce{N2 + H + M &-> HN2 + M},\\
	\textbf{\ce{HN2 + CO2}} &\textbf{\ce{-> HNCO + NO}}\label{eq:NH3_limiting_C_a},\\
	\ce{NO + H2 + M &-> NH2O + M},\\
	\textbf{\ce{NH2O + CHO}} &\textbf{\ce{-> NH2OH + CO}}\label{eq:NH3_limiting_C_b},\\
	\ce{NH2OH &-> NH2 + HO},\\
	\ce{NH2 + H2 &-> NH3 + H},\\
	\ce{CO + H + M &-> CHO + M},\\
	\ce{HO + H2 &-> H2O + H}.\\
	{\rm Net:} \; \ce{N2 + CO2 + 3\,H2 &-> NH3 + HNCO + H2O}.
\end{align}
\end{linenomath*}
Pathway C is limited by reaction \eqref{eq:NH3_limiting_C_a} at 2000\,K for $f\mathrm{O}_2$=IW-3, and limited by reaction \eqref{eq:NH3_limiting_C_b} for $f\mathrm{O}_2$=IW-2 and above. The pathway is limited by \eqref{eq:NH3_limiting_C_b} at lower temperatures.

Pathway D:
\begin{linenomath*}
\begin{align}
	\ce{N2 + H + M &-> HN2 + M},\\
	\ce{HN2 + CO2 &-> HNCO + NO},\\
	\textbf{\ce{NO + H + M}} &\textbf{\ce{-> HNO + M}}\label{eq:NH3_limiting_D_b},\\
	\ce{HNO + H2 &-> HN + H2O},\\
	\textbf{\ce{HN + CH4 + M}} &\textbf{\ce{-> CH5N + M}}\label{eq:NH3_limiting_D_a},\\
	\textbf{\ce{CH5N}} &\textbf{\ce{-> NH2 + CH3}}\label{eq:NH3_limiting_D_c},\\
	\ce{NH2 + H2 &-> NH3 + H},\\
	\ce{CH3 + H2O &-> CH4 + HO},\\
	\ce{HO + H2 &-> H2O + H}.\\
	{\rm Net:} \; \ce{N2 + CO2 + 3\,H2 &-> NH3 + HNCO + H2O}.
\end{align}
\end{linenomath*}
Pathway D is limited by reaction \eqref{eq:NH3_limiting_D_a} at 1500\,K and above. At 1500\,K the pathway is limited by reaction \eqref{eq:NH3_limiting_D_a} at $f\mathrm{O}_2$=IW+2 and above, and is limited by reaction \eqref{eq:NH3_limiting_D_b} and $f\mathrm{O}_2$=IW+1 and below. At 800\,K the pathway is limited by \eqref{eq:NH3_limiting_D_b} at $f\mathrm{O}_2$=IW+2 and below, and is limited by reaction \eqref{eq:NH3_limiting_D_c} at $f\mathrm{O}_2$=IW+3.5. This pathway is only significant in the conversion at 1000\,K, where it is subdominant and comparative with pathways B and D, because of the increased \ce{CH4} abundance at 1000\,K. This reflects how only the warm atmospheres are capable of producing significant abundances of \ce{CH4}, but not the cool atmospheres due to kinetic limitations to achieving equilibrium.

\begin{figure}[!htb]
	\begin{center}
	\includegraphics[width=0.95\textwidth]{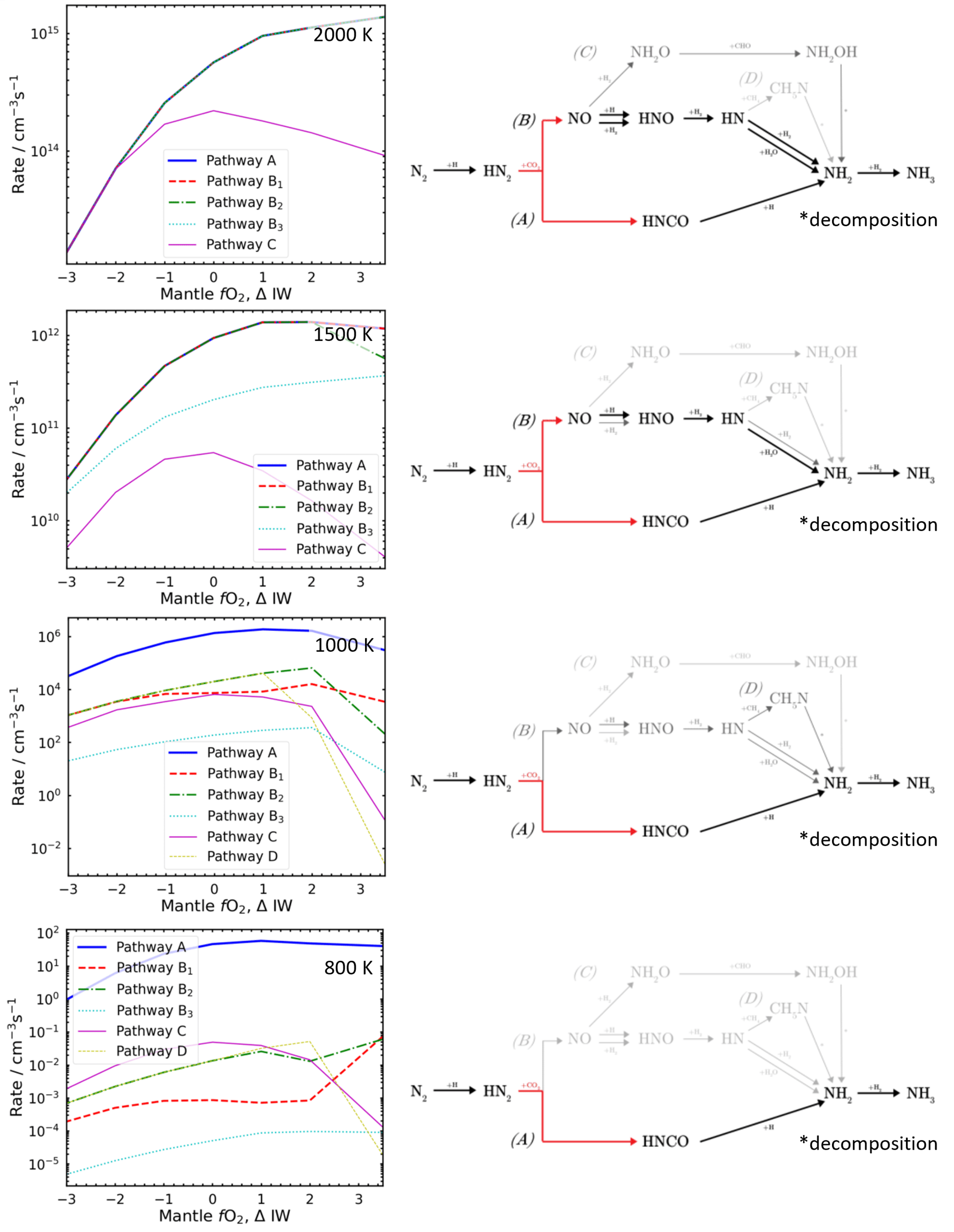}
	\caption{Reaction rates as a function of $f\mathrm{O}_2$ of the pathways that are dominant or subdominant at one or more of the temperatures investigated. Alongside we show the dominant pathway highlighted in the reaction network in bold, and its limiting reaction highlighted in red. From top to bottom, the figures correspond to atmospheric temperatures of 2000\,K, 1500\,K, 1000\,K, and 800\,K.}
	\label{fig:nh3_pathways}
\end{center}	
\end{figure}

At high temperature the branches via pathways A and B both dominate in the conversion of \ce{N2 -> NH3} thus a combined reaction pathway and net reaction are as follows:

\begin{linenomath*}
\begin{align}
        \ce{N2 + CO2 + 3\,H2 &-> NH3 + HNCO + H2O},\\
	\ce{N2 + H + M &-> HN2 + M},\\
	\ce{HN2 + CO2 &-> HNCO + NO},\\
	\ce{HNCO + H &-> NH2 + CO},\\
	\ce{NH2 + H2 &-> NH3 + H},\\
	\ce{CO + HO &-> CO2 + H},\\
	\ce{H2O + H &-> HO + H2},\\
	{\rm Net:} \; \ce{N2 + 3\,H2 &-> 2\,NH3}.
\end{align}
\end{linenomath*}

Compared to previously identified reaction schemes for the \ce{N2 -> NH3} conversion at high temperatures, the net reaction is the same however the mechanism of conversion is different. Previously identified reaction schemes for \ce{H2}-atmospheres have followed sequential reactions of \ce{N2} with \ce{H} and \ce{H2} that lead to a single bonded \ce{H_xN \bond{-} NH_y} species and the single bond is then broken leading to two \ce{NH_{x/y}} species that can then quickly form \ce{NH3} \cite{Moses2011,Line2011}. The dominant scheme here instead features the reaction \ce{HN2 + CO2 -> HNCO + NO} which breaks the \ce{N = N} double bond, and is responsible for both splitting the scheme into different branches and limiting the overall conversion to \ce{NH3}, with rate coefficient: $k = 3.32\times10^{-11}\rm{exp}(-12630\,K/T)$ \cite{Tomeczek2003}. The reaction is endothermic, with the numerator of the exponent determined by the Gibbs free energy which varies with temperature. The chemical-kinetics simulation uses the temperature dependent value. As quoted here, it is evaluated at 900\,K, the approximate quench temperature that was found for \ce{NH3} in our analysis.

\clearpage

%
%



\bibliography{refs}

\end{document}